\documentclass[10pt,journal,compsoc]{IEEEtran}
\usepackage{cite}
\usepackage{amsmath,amssymb,amsfonts}
\usepackage{algorithmic}
\usepackage{graphicx}
\usepackage{textcomp}
\usepackage{cite}
\usepackage{amsmath,amssymb,amsfonts}
\usepackage{algorithmic}
\usepackage{graphicx}
\usepackage{paralist}
\usepackage{caption}
\usepackage{subcaption}
\usepackage{amsmath}
\usepackage{soul}
\renewcommand\hl[1]{#1}
\usepackage{blkarray}
\usepackage{textcomp}
\usepackage{tabulary}
\usepackage{booktabs}

\begin{document}


\title{Scaling and Placing Distributed Services on Vehicle Clusters in Urban Environments}

\author{Kanika Sharma,~\IEEEmembership{}
        Bernard Butler,~\IEEEmembership{}
        and~Brendan~Jennings,~\IEEEmembership{}
\IEEEcompsocitemizethanks{\IEEEcompsocthanksitem K. Sharma, B. Butler and B. Jennings are with the Walton Institute, Waterford Institute of Technology, Ireland.\protect\\
E-mail: \{kanika\_sharma,bbutler,bjennings\}@ieee.org
}}

\IEEEtitleabstractindextext{

\begin{abstract}

  \hl{Many vehicles spend a significant amount of time in urban traffic congestion. Due to the evolution of autonomous cars, driver assistance systems, and in-vehicle entertainment, many vehicles have plentiful computational and communication capacity. How can we deploy data collection and processing tasks on these (slowly) moving vehicles to productively use any spare resources? To answer this question, we study the efficient placement of distributed services on a moving vehicle cluster. We present a macroscopic flow model for an intersection in Dublin, Ireland, using real vehicle density data. We show that such aggregate flows are highly predictable (even though the paths of individual vehicles are not known in advance), making it viable to deploy services harnessing vehicles' sensing capabilities. Our main contribution is a detailed mathematical specification for a task-based, distributed service placement model that scales according to the resource requirements and is robust to the changes caused by the mobility of the cluster. We formulate this as a constrained optimization problem, with the objective of minimizing overall processing and communication costs. Our results show that jointly scaling tasks and finding a mobility-aware, optimal placement results in reduced processing and communication costs compared to an autonomous vehicular edge computing-based na\"{i}ve solution.}
\end{abstract}
\begin{IEEEkeywords}
Fog Computing, Vehicular Cloud Computing, Vehicular Fog Computing, Intelligent Transport Systems, Flexible service models, Internet of Things, Service Placement, Resource Allocation.
\end{IEEEkeywords}
}
\maketitle


%


\section{Introduction}
\label{sec:introduction}

Vehicles will be one of the most important agents in the emerging Internet of Things (IoT) ecosystem, owing to their embedded sensors and built-in cameras, which can be used to capture contextual data for object detection and surveillance \cite{8641431}. Since each vehicle generates an average of 30 Tb of data per day, it is infeasible to send all the generated data to the Cloud using the controlled and limited cellular bandwidth \cite{8114558}. The increasing number of Smart vehicles and overall vehicular traffic has inspired the concept of Vehicular Fog Computing (VFC) \cite{9097892,7415983}, where vehicles are utilised as Fog nodes and play the role of service providers. This new data generation and communication paradigms is motivated by Fog Computing \cite{8641431} and Mobile Edge computing based models \cite{8637930,8786080} which provide ubiquitous connectivity and location-aware network responses at the edge of the network, complemented with cloud computing in the network core.

Closely-spaced, moving vehicles, could be used to, for example, monitor the compliance of both vehicles and pedestrians to lock-down restrictions introduced in response to the COVID-19 pandemic, by capturing data from essential service vehicles. The video data can also be collected for estimating usage patterns of highways for urban planning, reducing the need for installing more roadside infrastructure. Slowly moving vehicles can also be used to collect 3D roadmap data, to increase the perception range of intelligent vehicles, reducing the need for sending high definition data to the Cloud \cite{8946549},\cite{8950168}. All these use cases require rich computation and communication resources so that this data can be used for insights and decision-making. 

In this paper, we propose that vehicles lease their otherwise unused processing, communication, and storage resources to collaboratively host data analytics services that can pre-process and filter the data they collect. Thus, a dense group (or cluster) of moving vehicles can cooperatively execute distributed services that comprise: (i) delay-sensitive tasks that have a short sense-actuate cycle and require real-time decision making, and/or (ii) data collection and analytics tasks that are location- and context-specific.

\begin{figure}
\centering
\begin{subfigure}[b]{0.55\textwidth}
   \includegraphics[width=0.95\linewidth,height=6cm]{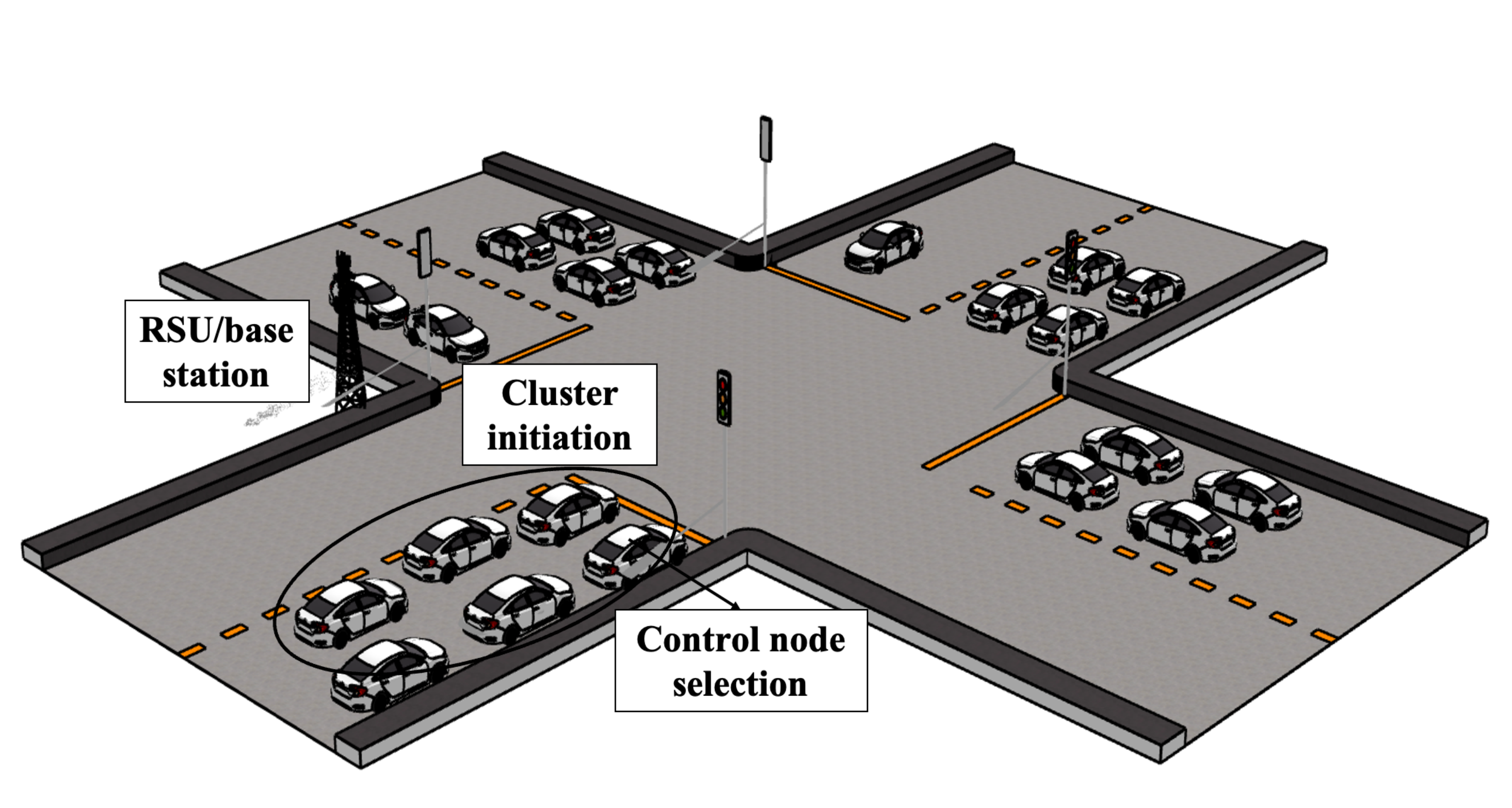}
   \caption{Vehicle cluster at state $t_1$}
   \label{fig:Ng1} 
\end{subfigure}

\begin{subfigure}[b]{0.55\textwidth}
   \includegraphics[width=0.95\linewidth,height=6cm]{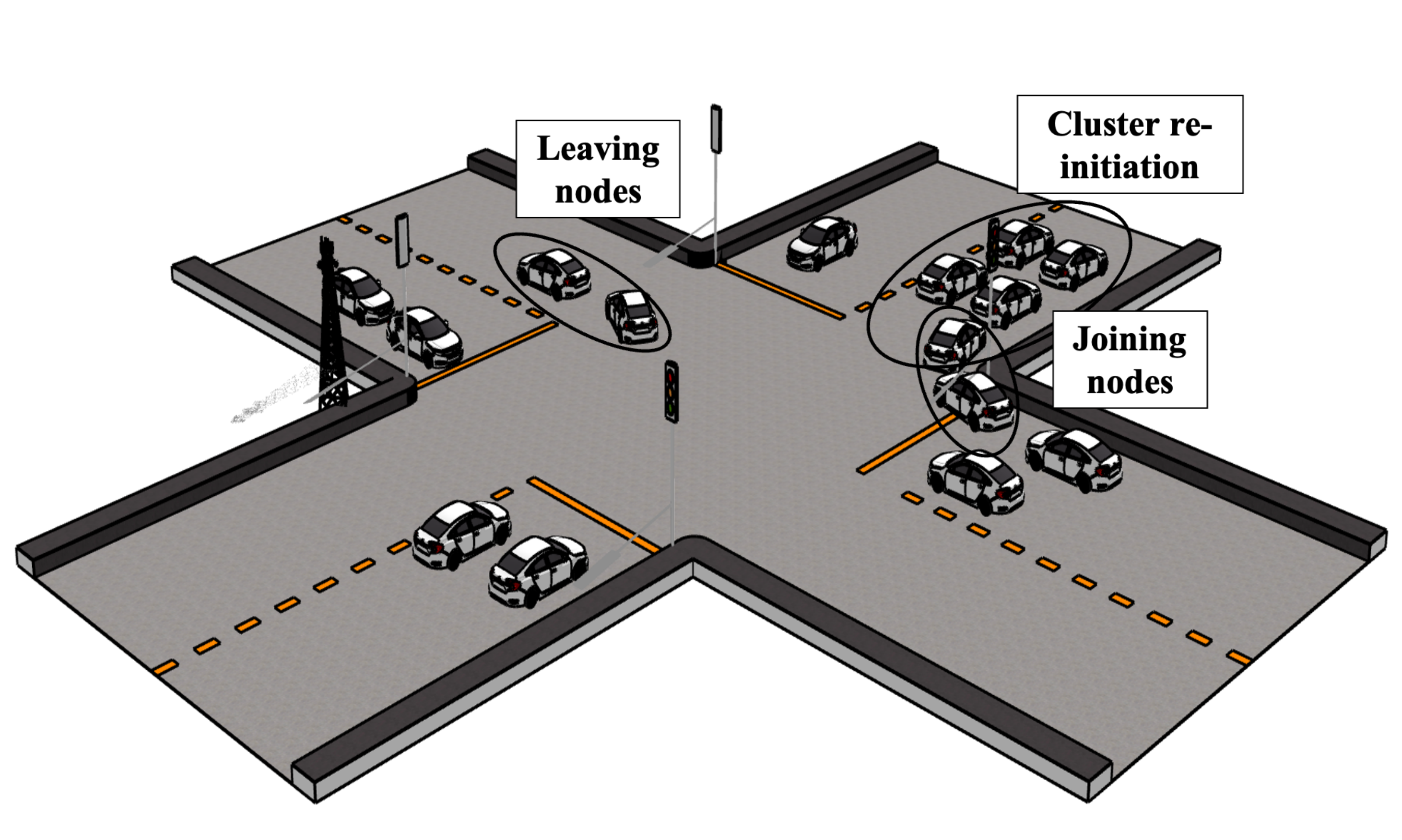}
   \caption{Vehicle cluster at state $t_2$}
   \label{fig:Ng2}
\end{subfigure}

\caption{Vehicle Clusters form, but membership changes over time. Clusters accept service placement requests from RSUs and perform scaling and placement of the accepted service. Fig (a) and (b) depict the state of the cluster over time $t_1$ and $t_2$. The mobility of the vehicles requires cluster re-initiation.}
\label{fig:scenario}
\end{figure}

Vehicles are distinguished through their \emph{mobility} (in particular, vehicles join and leave a cluster in a stochastic manner), so the resource allocation task becomes time-dependent. Mobility affects both network connectivity and computation capacity, and hence the Quality-of-Service (QoS). Our work aims to utilize the aggregate mobility behavior of vehicles to select reliable vehicle nodes, i.e., those that have a higher probability of staying with a given cluster of vehicles, in order to avoid service failure and reduce the need for service reconfiguration.  As depicted in Fig.~\ref{fig:scenario},  one vehicle, or Roadside Unit (RSU), acts as a managing entity to collect and update both the resource and mobility states of the cluster and enable flexible service scaling. 

We take the specific case of using in-built cameras in cars that are willing to lease their resources, to provide streaming data on request. This data is processed by streaming it to linear chains of \hl{tasks}, where each task has different processing functionality. \hl{One linear chain of tasks form a service that satisfies a service request.}  We employ a component-oriented distributed service model, where each \hl{task} can be realised via a collection of multiple \hl{task instances (TIs)}; in this manner each \hl{task} can be scaled out according to the demand and the available infrastructure resources. For example, using multiple camera instances increases the spatio-temporal coverage of the data collection, thereby increasing the scope for more accurate and efficient data analysis, especially in applications like building 3D road maps for self-driving cars. Moreover, having replicas of computing tasks enables the utilization of distributed resources and reduces the impact of nodes leaving the vehicle cluster. Node and link failure in this \hl{service} model requires only the replication of a problematic \hl{TI} onto a more suitable vehicle so the \hl{service} chain still works, instead of re-configuring the entire service. As the \hl{tasks} are data-dependent, even if the cluster is computationally rich, it needs to be split into smaller \hl{TIs} if the link capacity between host nodes is not enough. The \hl{task} placement also needs to avoid placing \hl{TIs} on those resource-rich nodes that have a high probability of leaving the cluster. Thus, \hl{service} deployments can be adapted at run-time according to both the known resource availability and the predicted mobility state of the cluster. 

This paper builds upon our previous work \cite{8581027}, where we introduced the problem of the placement of distributed data collection services on a moving vehicle cluster. We modeled the service placement problem as an optimization problem subject to constraints related to node resource capacity, link capacity, distributed application deployment (full deployment, anti-collocation and adjacency constraints) and vehicle mobility. In this paper the following novel contributions are made:
\begin{compactitem}

    \item we introduce a macroscopic flow model for vehicle mobility using real vehicle density data from  traffic data collected in Dublin, Ireland. We highlight the predictability in vehicle traffic for different time intervals, using Multivariate Linear Regression with an accuracy of 93 to 99\%. 
    
    \item we formulate the \hl{service placement} problem mathematically in two parts: 1) a flexible and distributed service model with data-dependent \hl{tasks} instead of \hl{static service templates}; and 2) a mobility-aware infrastructure model;
    \item to validate our approach, we simultaneously place \emph{two} services with diverse resource requirements (either computation or communication intensive). We also impose multi-tenancy constraints to reuse the same nodes and/or \hl{TIs} for different applications;
    \item we optimize the service scaling and placement problem by first scaling the linear chain at run-time and then finding a placement plan that optimizes resource usage over a multi-hop cluster, repeating as necessary.  
    \item we compare our approach to the na\"{i}ve solution, presented in \cite{7847322}. We also validate the applications on a Fog Computing simulator.
\end{compactitem}

The paper is organized as follows: \S2 describes related research in the field of service placement and task offloading schemes in vehicular networks. In \S3, we describe the motivation behind using vehicles as infrastructure, estimate vehicular cluster capacity, and validate vehicle traffic predictability with Multivariate Linear Regression. In \S4, we define the system model and the network topology and in \S5, we define the constraints and the mathematical formulation of the placement problem. \hl{In \S6, we introduce application types and run an experiment for an application scenario on a Fog simulator. We then discuss solving the optimization problem, simulation setup and our results. Finally, in \S7 we conclude the work and give an outline of our future work.}


\section{Related Work} 

Gerla~et al.\ \cite{6257116} was the first to introduce the term Vehicular Cloud Computing (VCC) as a distributed computing platform, wherein vehicles form a micro cloud in an ad hoc manner. \hl{They identified many important applications for VCC, including urban sensing by uploading videos of congestion and pavement conditions that other vehicles could access.} Hou et al.~\cite{7415983} were the \hl{first to introduce the concept of Vehicular Fog Computing (VFC) as an architecture that can be used to enable multiple end-user or edge devices to collaborate to carry computation and communication tasks.} They considered both slow-moving and parked vehicles and analyzed the quantitative capacity of such a Vehicular Fog. Ma et al.~\cite{8756836} introduced a Platoon-assisted Vehicular Edge Computing system based on the stability of the platoon in vehicular networks. \hl{They were the first to introduce a Reinforcement Learning (RL)-based optimization scheme to obtain optimal price strategy of task flows. Lee et al.~\cite{9097892} also modified an RL-based algorithm to make efficient resource allocation decisions leveraging vehicles' movement and parking status to minimise service latency.}


Zhao et al.~\cite{8745530} jointly optimize the computation offloading decision and computation resource allocation in vehicular networks. They designed a collaborative optimization scheme where offloading decisions are made through a game-theoretic approach and resource allocation is achieved using the Lagrange multiplier method. The feasibility of using vehicles as Fog nodes for video crowd-sourcing and real-time analytics has been studied by Zhu et al.~\cite{8493119}. They evaluated the availability of client nodes that generate data in proportion to the vehicle Fog nodes that process the data, using processing capacity on on-board units. \hl{However, they focus solely on the data transmission problem in the model.} Xiao et al.~\cite{8756132} also evaluated the achievable performance of a vehicular cloud and analyze the total computation capacity for the same. They also model vehicle mobility patterns using parameters like staying time and the incoming and outgoing flow rate of vehicles. \hl{This capacity analysis is a crucial requirement for enabling a vehicular computing model but they do not focus on the service model and applications to be deployed.} In Kong et al.~\cite{8242676}, the traditional mobility models for vehicles are replaced by methods based on social patterns, community interest group check-ins on social media data etc. 

The mobility of vehicle nodes makes the task allocation problem more challenging in Vehicular Fog Computing. Zhu et al.~\cite{8489874} introduced an event-driven dynamic task allocation framework designed to reduce average service latency and overall quality loss. Both multi-source data acquisition and distributed computing in Fog-computing-based intelligent vehicular network are studied by Zhang et al.~\cite{8198803}. \hl{They introduce a hierarchical, QoS-aware resource management architecture, but consider the Fog servers as static.} Vehicular micro cloud has been studied as virtual edge servers for efficient connection between cars and back-end infrastructure in Hagenauer et al.~\cite{Hagenauer:2017:VMC:3131944.3133937}. They use map-based clustering at intersections, as intersections have line of sight in multiple directions which result in better connectivity between the Cluster Heads (CHs) and other cluster members. \hl{Even though they primarily focus on cluster creation and cluster head selection, they evaluate a data collection application, with varying data aggregation rates at the CH. Goudarzi et al.~\cite{8960404} introduced an application placement technique for concurrent IoT applications in Edge and Fog computing environments. They propose a a batch application placement technique based on the Memetic Algorithm to efficiently place tasks of different workflows on appropriate IoT devices, fog servers, or cloud servers.}

Fog computing based information-centric approaches have been studied by Wu et al.~\cite{8767078}, where Fog nodes act as network monitor for cognitive caching and computational resource configurations. These techniques are crucial for future internet applications like automated driving enabled vehicles. \hl{Wu et al.~\cite{7847322} study the dynamic management and configuration of heterogeneous consensus for differentiated and dynamic IoT services in a blockchain. The paper also highlights the importance of dynamically switching heterogeneous consensuses of different applications, as the IoT node is re-utilized by many applications. The authors give an example of using deployed cameras on the roadside for both Intelligent Transport systems and for surveillance. This approach can be used for dynamic service requirements in a moving vehicle cluster.}



\section{Motivation}

Our work is motivated by the increasing number of \emph{Smart} and continuously connected cars, and the unresolved issue of vehicle congestion---especially in urban areas. Before introducing our service scaling and placement scheme we first provide justification that placing services on a vehicle cluster in order to provide time and/or location sensitive sensing functionality is a viable proposition. There are two important aspects: 1) whether traffic flows in an urban setting are likely to be predictable over the course of a day, and 2) whether a slow moving cluster will accommodate sufficient communications capacity between vehicles to facilitate service operation.

\subsection{Predictability of vehicle flows} \label{sec:flowmodel}

We find that vehicular flow in urban traffic zones is predictable throughout the day. We also show that the vehicular density pattern at an intersection follows a similar pattern of peak and off-peak flow through different weeks. We use macroscopic vehicle density data to create a generalised flow model for an intersection. This helps in classifying traffic flow into six different driving profiles. The vehicle clusters can then be initiated at the predicted peak traffic times, on any of the traffic flows with an assured density flow. 


\begin{figure}[t!]
\begin{subfigure}{0.5\linewidth}
  \includegraphics[width=\linewidth,height=4cm]{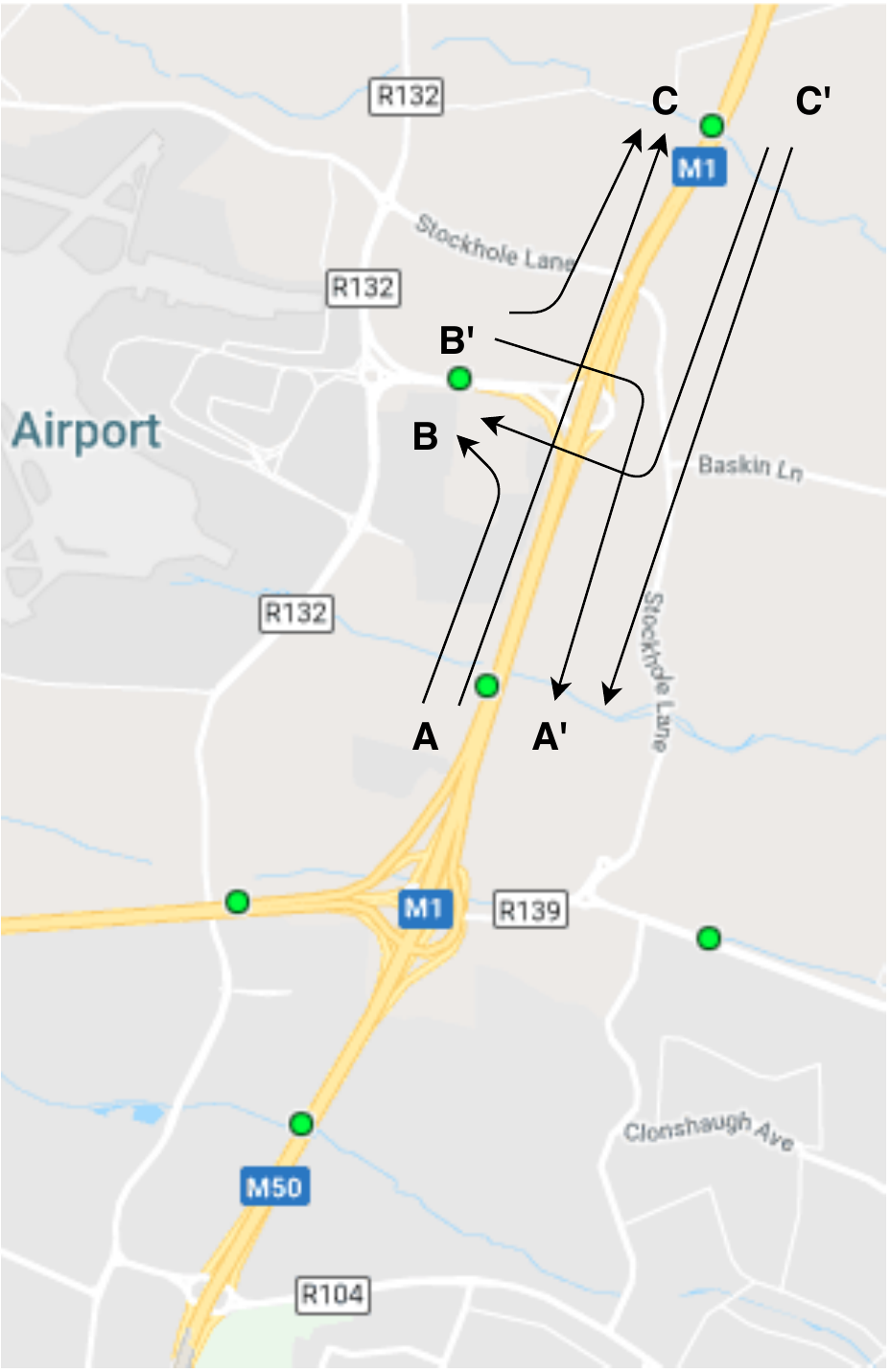}
 
\end{subfigure}\hfil 
\begin{subfigure}{0.5\linewidth}
  \includegraphics[width=\linewidth,height=4cm]{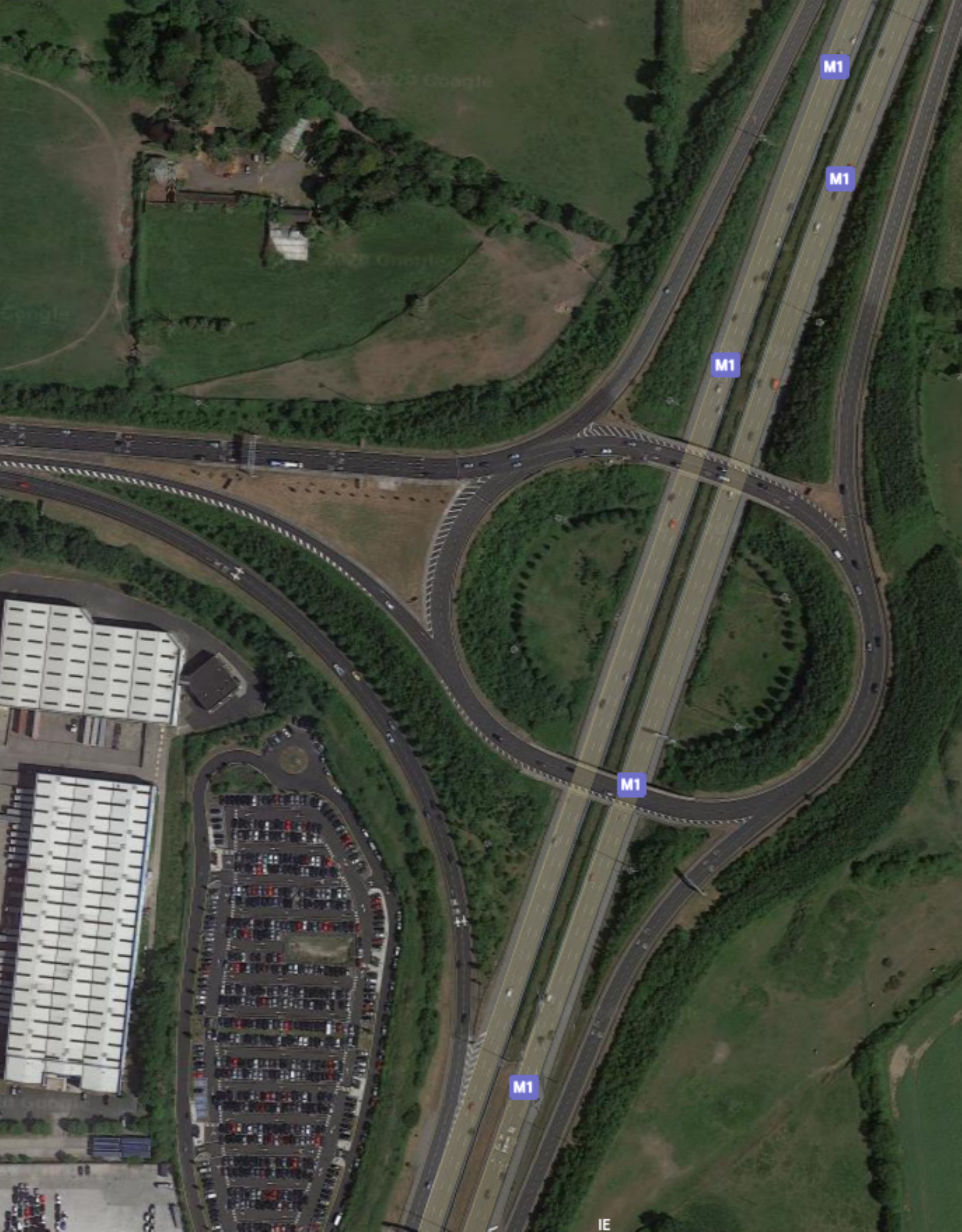}
\end{subfigure}\hfil 
\caption{Flow model at the selected Dublin intersection with six traffic flows from A to B, A to C, C' to A', C' to B, B' to C and B' to A}
\label{fig:flowmodel}
\end{figure}

\begin{figure}[t!]
\begin{subfigure}{0.48\linewidth}
  \includegraphics[width=\linewidth,height=3cm]{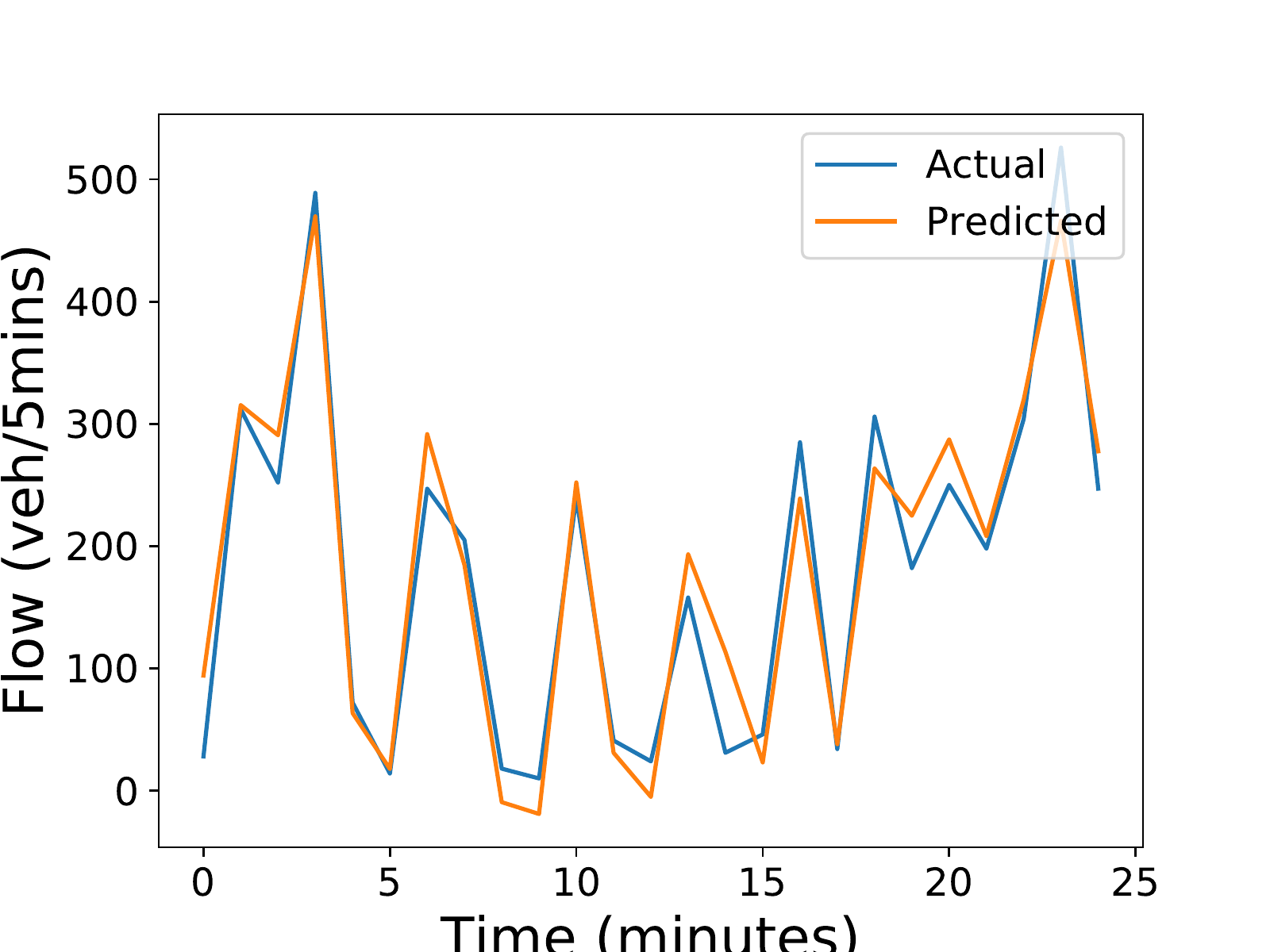}
  \caption{Linear regression model for traffic prediction every 5 minutes}
  \label{fig:LR5mins}
  \end{subfigure}\hfil 
\begin{subfigure}{0.48\linewidth}
  \includegraphics[width=\linewidth,height=3cm]{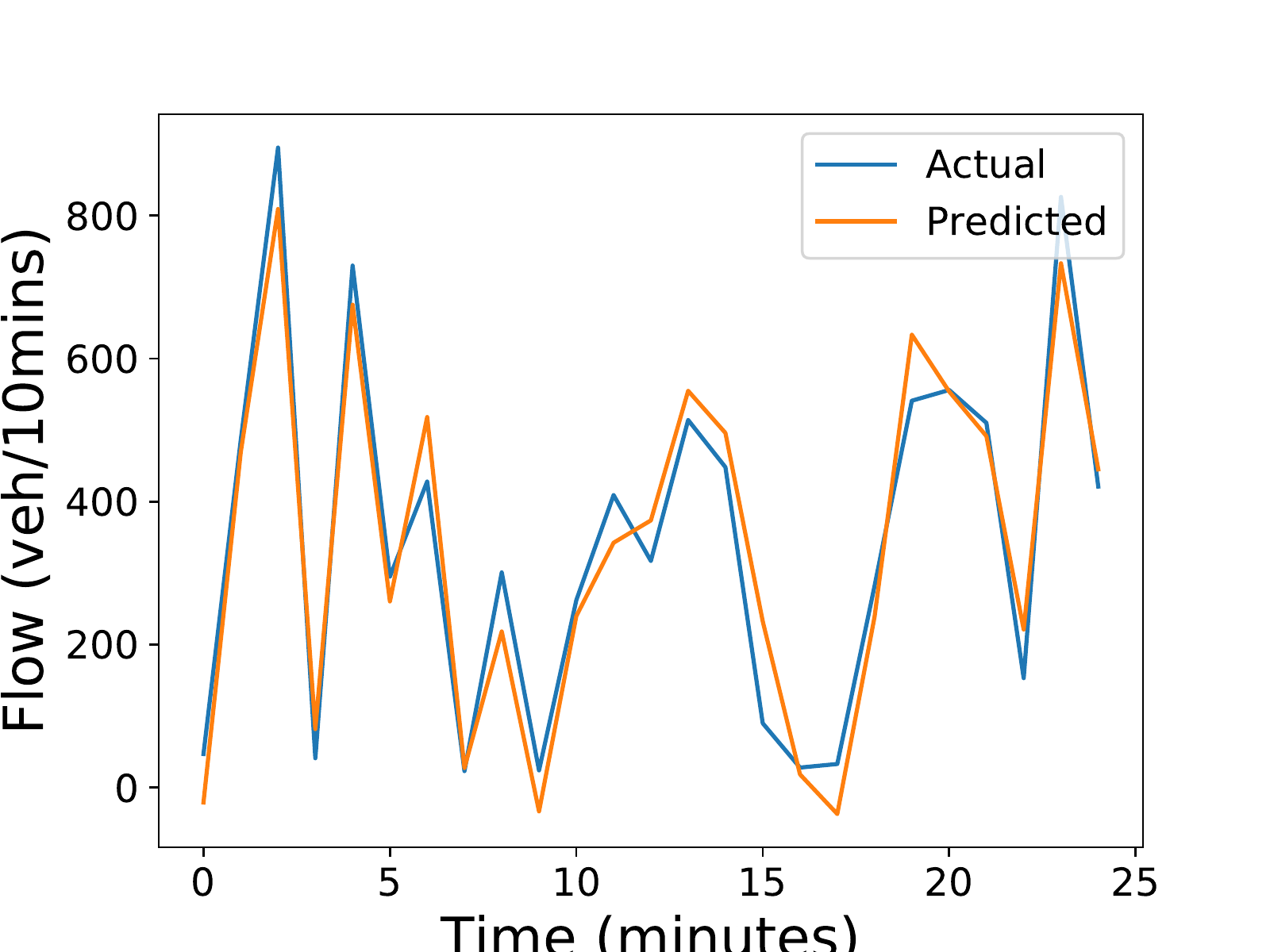}
  \caption{Linear regression model for traffic prediction every 10 minutes}
  \label{fig:LR10mins}
\end{subfigure}\hfil 

\begin{subfigure}{0.48\linewidth}
  \includegraphics[width=\linewidth,height=3cm]{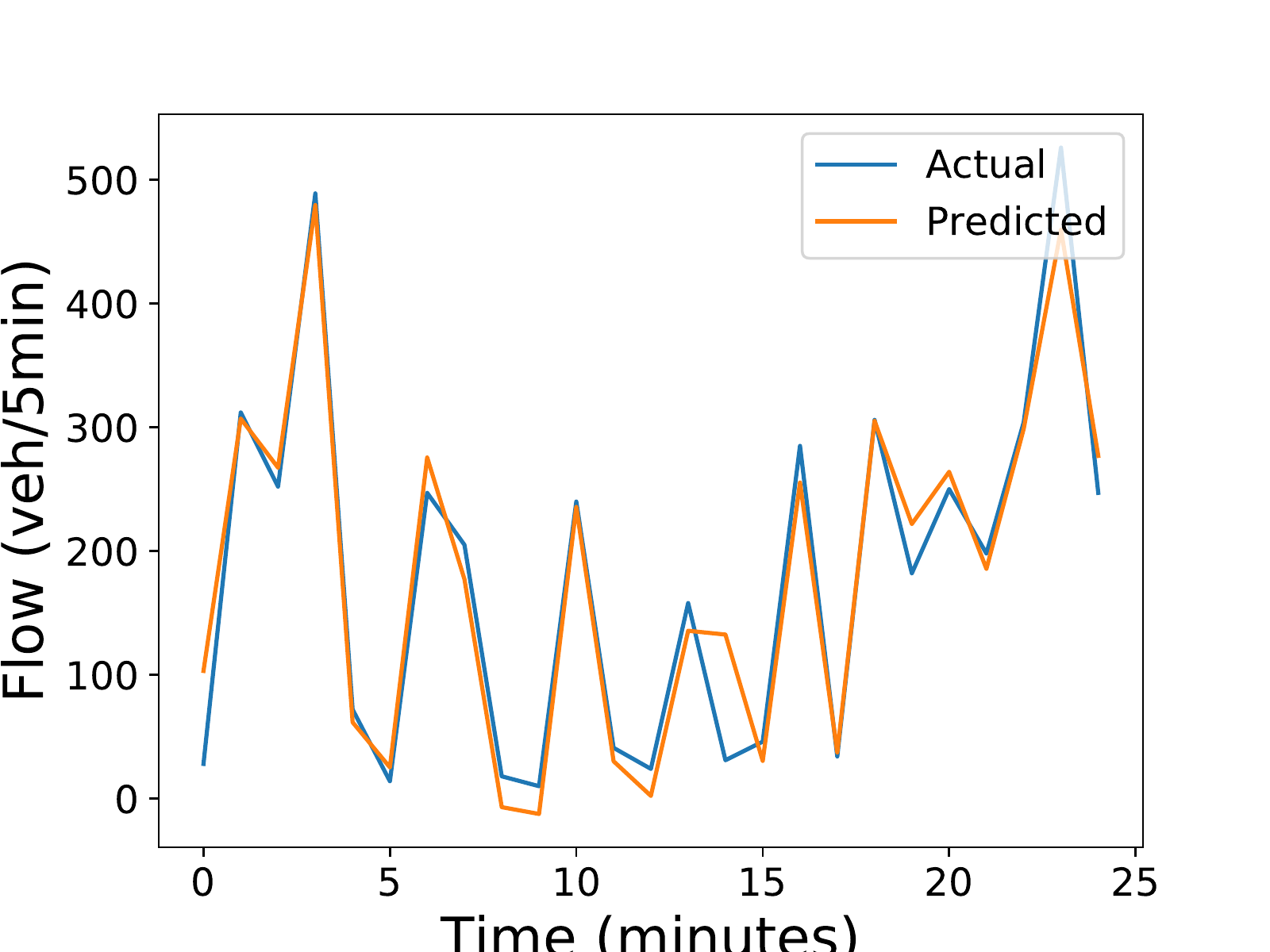}
  \caption{Multivariate linear regression model for traffic prediction every 5 minutes}
  \label{fig:MV5min}
\end{subfigure}\hfil 
\begin{subfigure}{0.48\linewidth}
  \includegraphics[width=\linewidth,height=3cm]{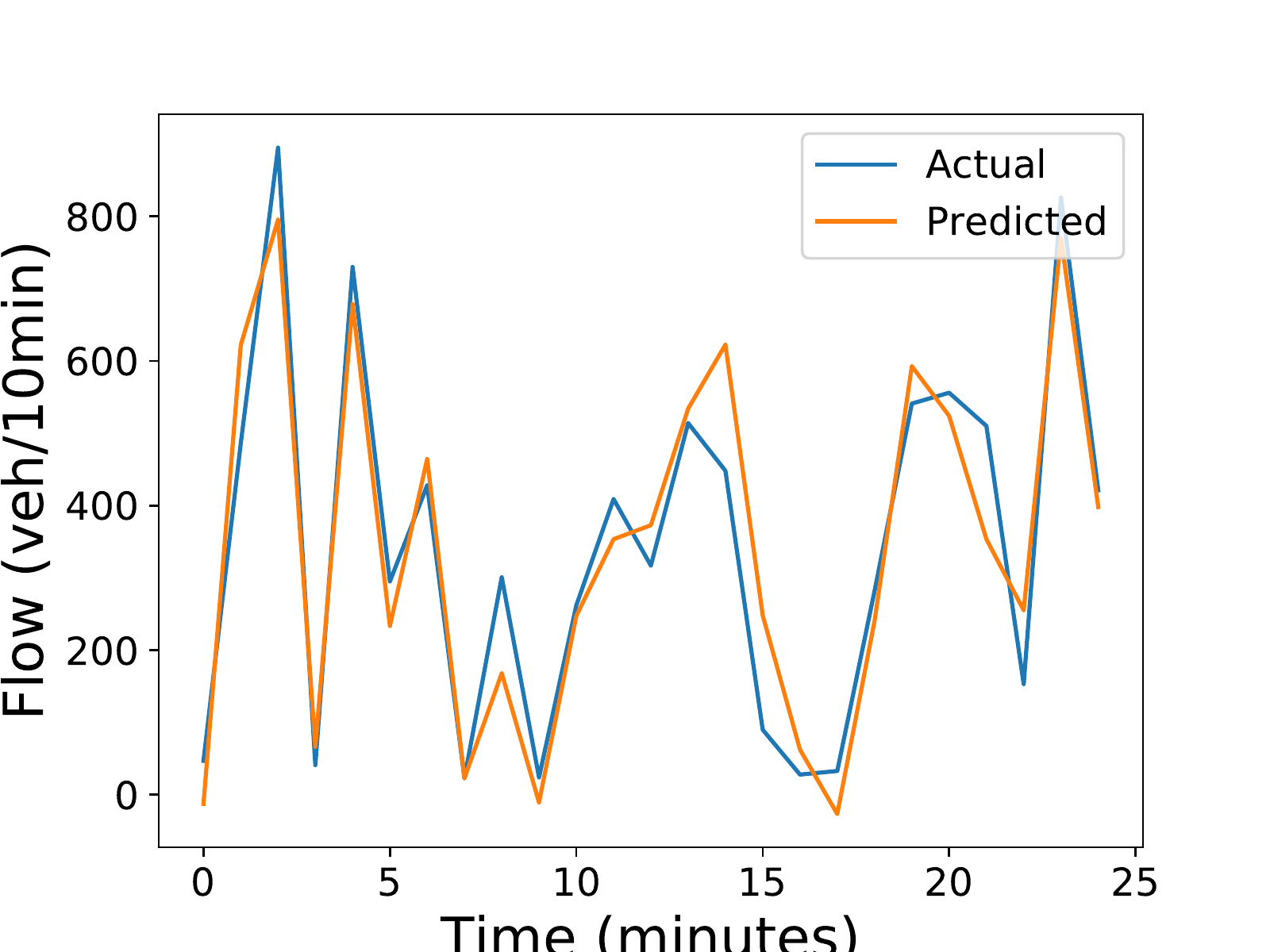}
  \caption{Multivariate linear regression model for traffic prediction every 10 minutes}
  \label{fig:MV10min}
\end{subfigure}\hfil 
\begin{subfigure}{0.48\linewidth}
  \includegraphics[width=\linewidth,height=3cm]{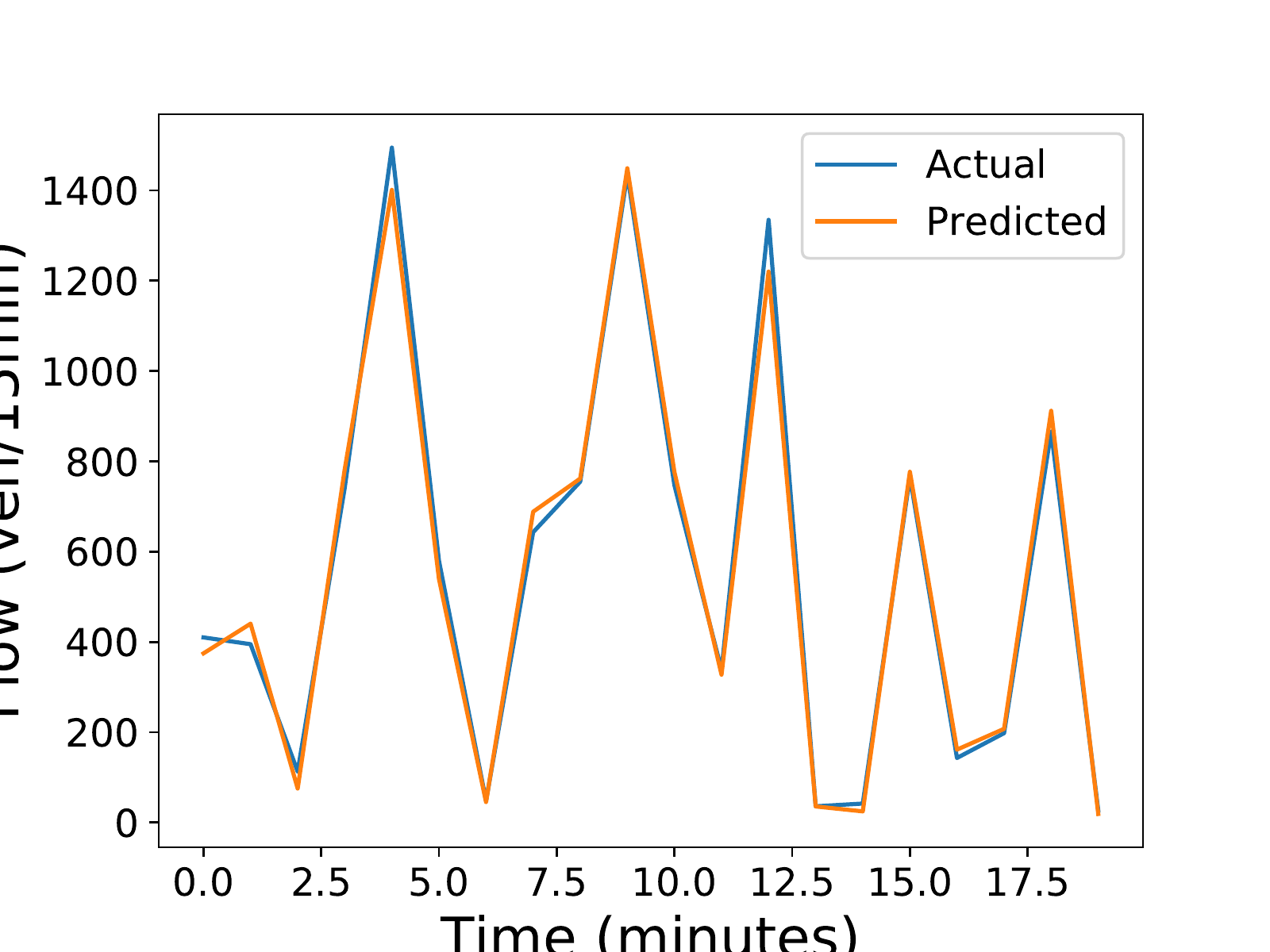}
  \caption{Multivariate linear regression model for traffic prediction every 15 minutes}
  \label{fig:MV15min}
\end{subfigure}\hfil
\begin{subfigure}{0.48\linewidth}
  \includegraphics[width=\linewidth,height=3cm]{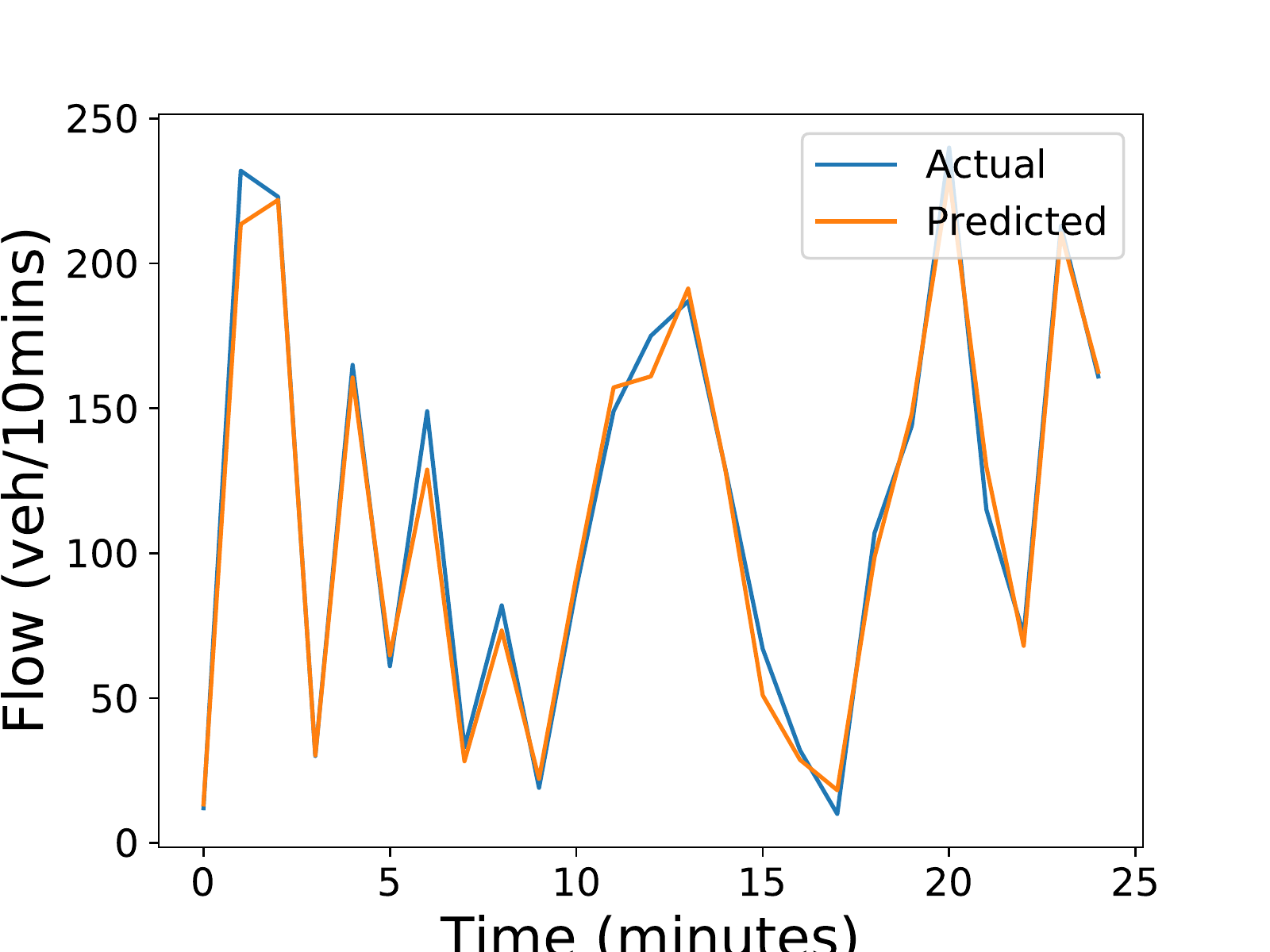}
  \caption{Multivariate linear regression model for traffic prediction every 10 minutes, April 2020}
  \label{fig:April10mins}
\end{subfigure}\hfil
\begin{subfigure}{0.48\linewidth}
  \includegraphics[width=\linewidth,height=3cm]{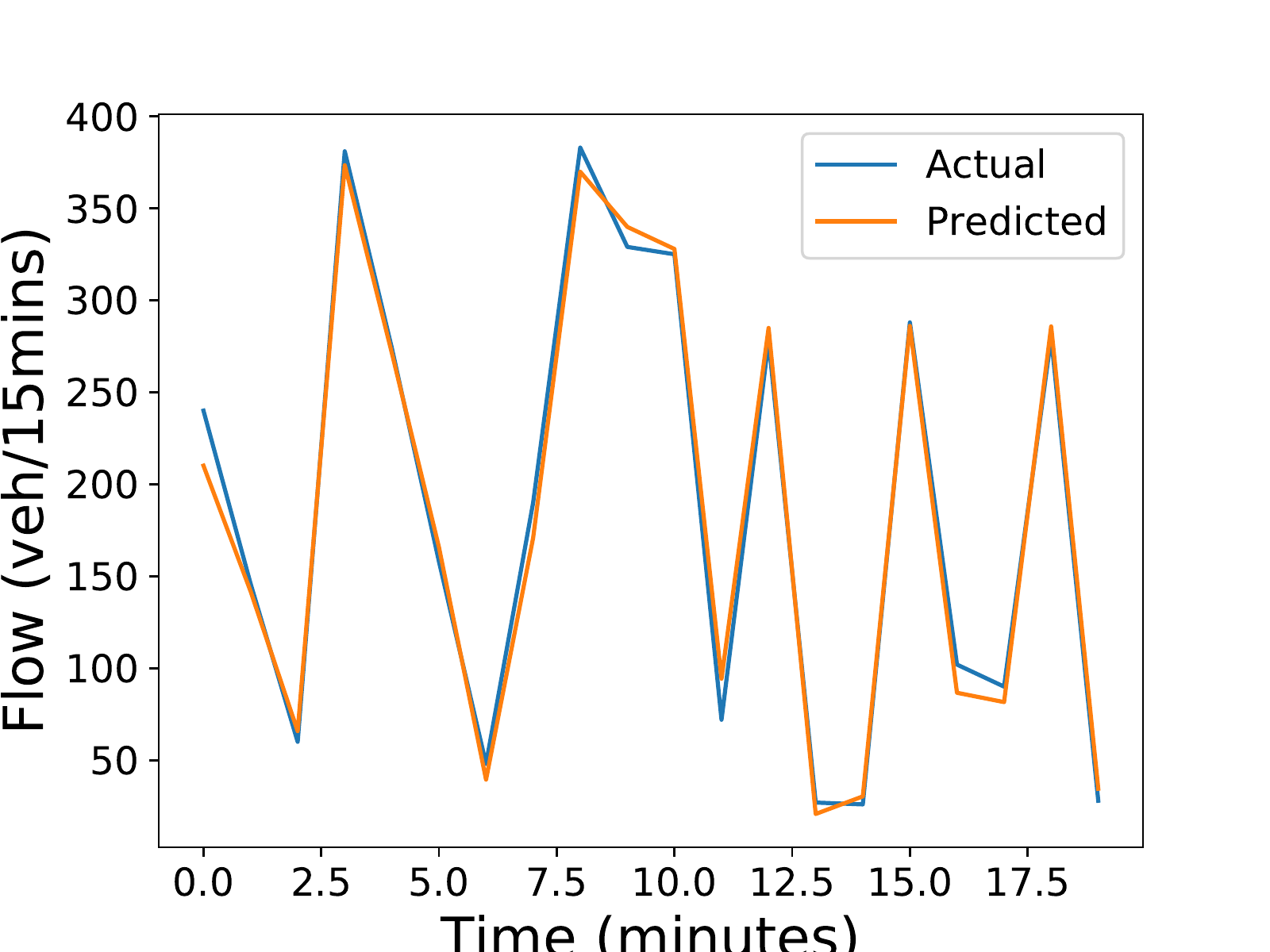}
  \caption{Multivariate linear regression model for traffic prediction every 15 minutes, April 2020}
  \label{fig:April15mins}
\end{subfigure}\hfil
\begin{subfigure}{0.48\linewidth}
  \includegraphics[width=\linewidth,height=3cm]{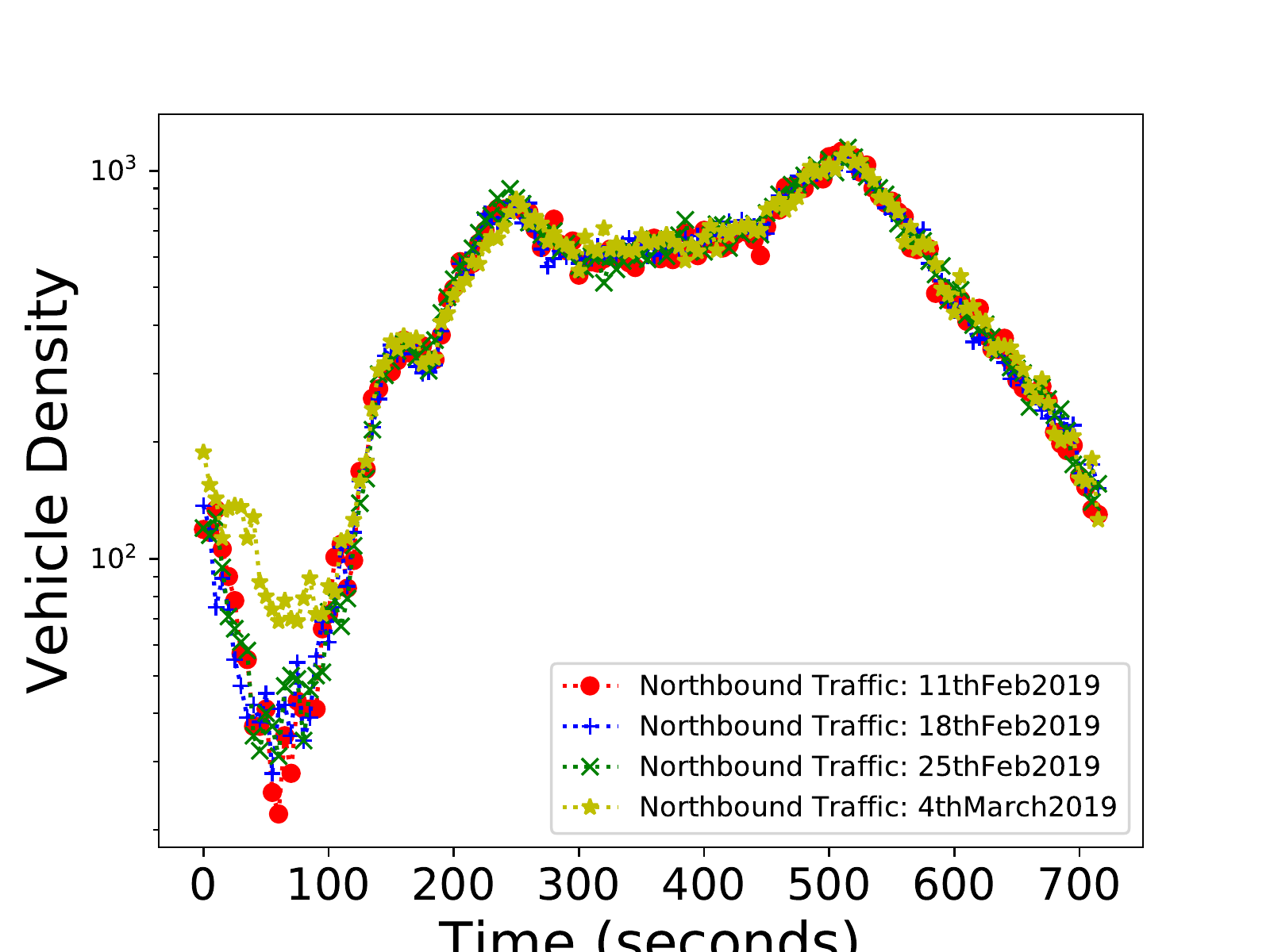}
  \caption{Vehicle density recorded every 10 minutes}
  \label{fig:5mins}
\end{subfigure}\hfil 
\begin{subfigure}{0.48\linewidth}
  \includegraphics[width=\linewidth,height=3cm]{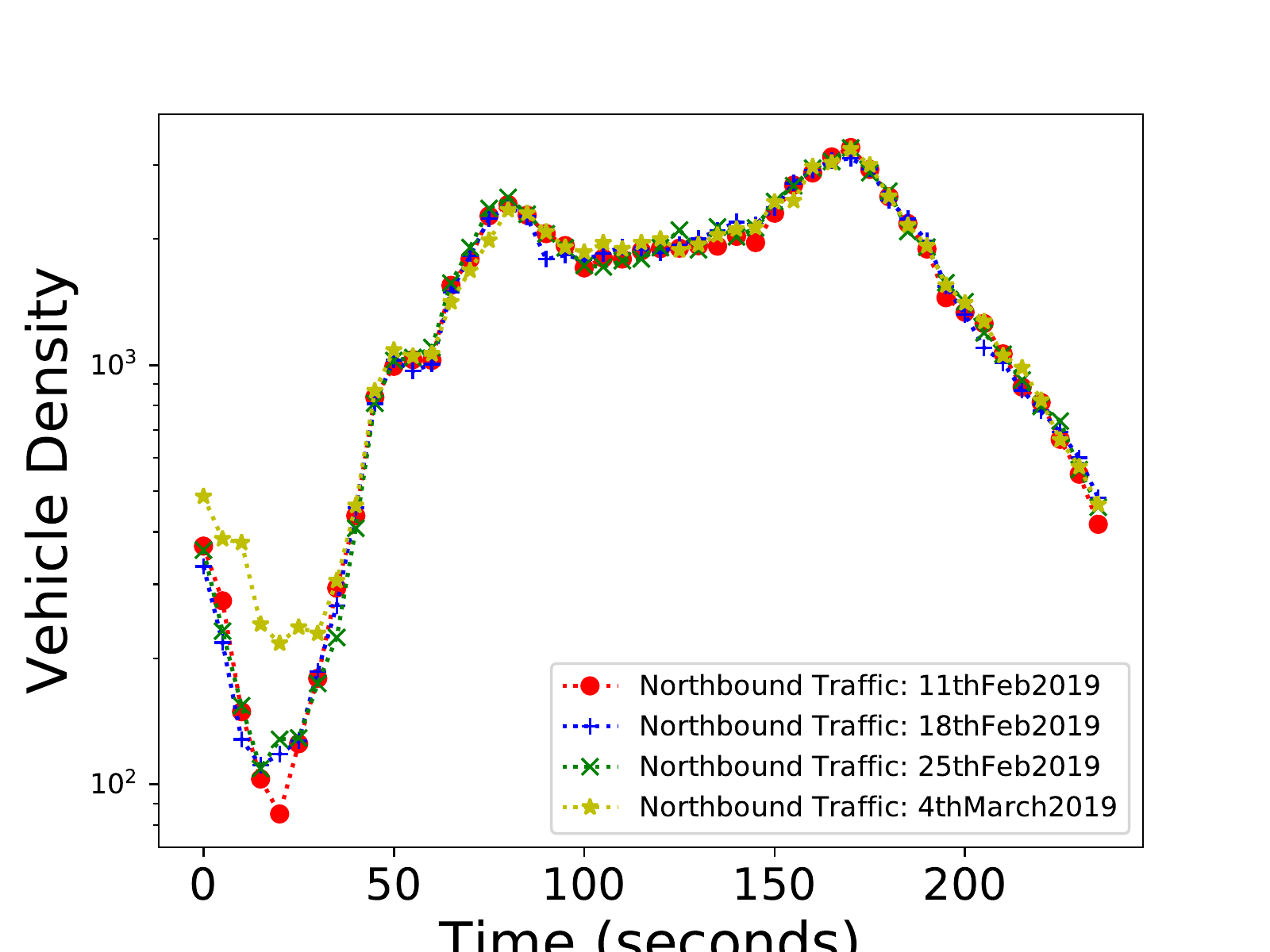}
  \caption{Vehicle density recorded every 30 minutes}
  \label{fig:30mins}
\end{subfigure}\hfil 
\caption{Traffic Prediction using real vehicle density data; these data depict the consistent and predictable vehicle densities at the intersection for all of the four weeks analyzed.}
\end{figure}

We first focus on a road network near Dublin Airport, using the vehicle flow data captured by the Transport Infrastructure Ireland Traffic Data website \footnote{https://www.nratrafficdata.ie, available: 22/10/20.}. A vehicular flow is defined as the number of detected vehicles passing a point in a period of time. The idea is to use the stochastic traffic flows at an intersection to predict the trajectory of a vehicle cluster. As depicted in (Fig.~\ref{fig:flowmodel}), we consider northbound flow from A to B and A to C, southbound flow from C' to A' and C' to B, eastbound traffic from B' to C and B' to A. We then employ a linear regression model to predict the traffic flow from one segment to the other, for all the six flows at the intersection. To understand the predictability of the traffic flows, we use the vehicle flow data, collected in the interval of 5, 10 and 15 minutes (based on the estimated travel time between any of the six points at peak and off-peak traffic time of the day) for a period of 24 hours. This data is used to model a generalized traffic flow model for an intersection.  We predict the vehicle density at point B taking into consideration the vehicle density at point A, using a linear regression model. We plot the actual and predicted incoming vehicle density at point B, for an interval of 5 minutes (\mbox{Fig.~\ref{fig:LR5mins}}) and 10 minutes  (\mbox{Fig.~\ref{fig:LR10mins}}).  The accuracy score of the prediction was 0.915 for a period of 5 minutes and 0.945 for 10 minutes respectively. This way, vehicles can be clustered in six different driving profiles for service execution, corresponding to the above-mentioned six flows. Table \ref{tab:linearity} depicts the r-value, p-value and the standard error for all the six flows.


We then use the vehicle flow data for the last 7 consecutive Mondays to predict a single flow, from A to B, using Multivariate Linear regression for data collected at an interval of  5 (\mbox{Fig.~\ref{fig:MV5min}}), 10 (\mbox{Fig.~\ref{fig:MV10min}}) and 15 (\mbox{Fig.~\ref{fig:MV15min}}) minutes. The same days in the week were studied to have similar patterns of mobility, within a range of a month to two, hence data for 7 consecutive Mondays was used. The predicted and actual vehicle flow at point B is depicted in Fig.~\ref{fig:MV5min}, ~\ref{fig:MV10min} and~\ref{fig:MV15min}. The $R^2$ accuracy score of the prediction was 0.93750, 0.9476 and 0.99216 for 5, 10 and 15 minutes respectively. We also considered the vehicle flow data during the period of COVID-19 lock-down, from 1st to 8th April 2020, to analyze the pattern of flow during the Coronavirus restrictions in Ireland. The restrictions resulted in much less traffic density at the intersection. \mbox{Fig.~\ref{fig:April10mins}} and \mbox{Fig.~\ref{fig:April15mins}} depict predicted vehicle flow using Multivariate Linear Regression, considering seven consecutive days during the lock-down, with an accuracy score of 0.97987 and 0.98746. We also plot the overall vehicular flow data for four consecutive Mondays, recorded in an interval of 10 minutes (\mbox{Fig.~\ref{fig:5mins}}) and 30 minutes (\mbox{Fig.~\ref{fig:30mins}}). The figures  depict the consistent and predictable vehicle density data for both northbound and southbound traffic for all four weeks. 


\begin{table}[tb]
\caption{r-value, p-value and standard error for predictability of the six flows at the intersection}
\label{tab:linearity}
\centering
\resizebox{\columnwidth}{!}{%
\begin{tabular}{cccccc}
\hline
 & \emph{Slope} & \emph{Intercept} & \emph{r value} & \emph{p value} & \emph{Standard error} \\ \hline
A -\textgreater B & 0.2799 & 75.1370 & 0.80550 & 4.4465 & 0.01728 \\ 
B’ -\textgreater C & 2.3044 & -86.3552 & 0.8487 & 4.0024 & 0.1204 \\ 
C’ -\textgreater A’ & 1.028840 & 100.13687& 0.97474& 2.45224 & 00.01978 \\
B’ -\textgreater A’ & .01091 & 64.3769 & 0.7500 & 2.8319 & 0.1488 \\ 
C’ -\textgreater B & 0.3786 & 85.0214 & 0.8706 & 1.3820 & 0.0179 \\ 
A -\textgreater C & 1.4932 & -27.3537 & 0.9604 & 1.0146 & 0.0363 \\ \hline
\end{tabular}
}
\end{table}

\begin{figure}[htbp]
\begin{minipage}[t]{\linewidth}

    

          \begin{center}
    \includegraphics[width=1.3\linewidth,height=5cm,keepaspectratio]{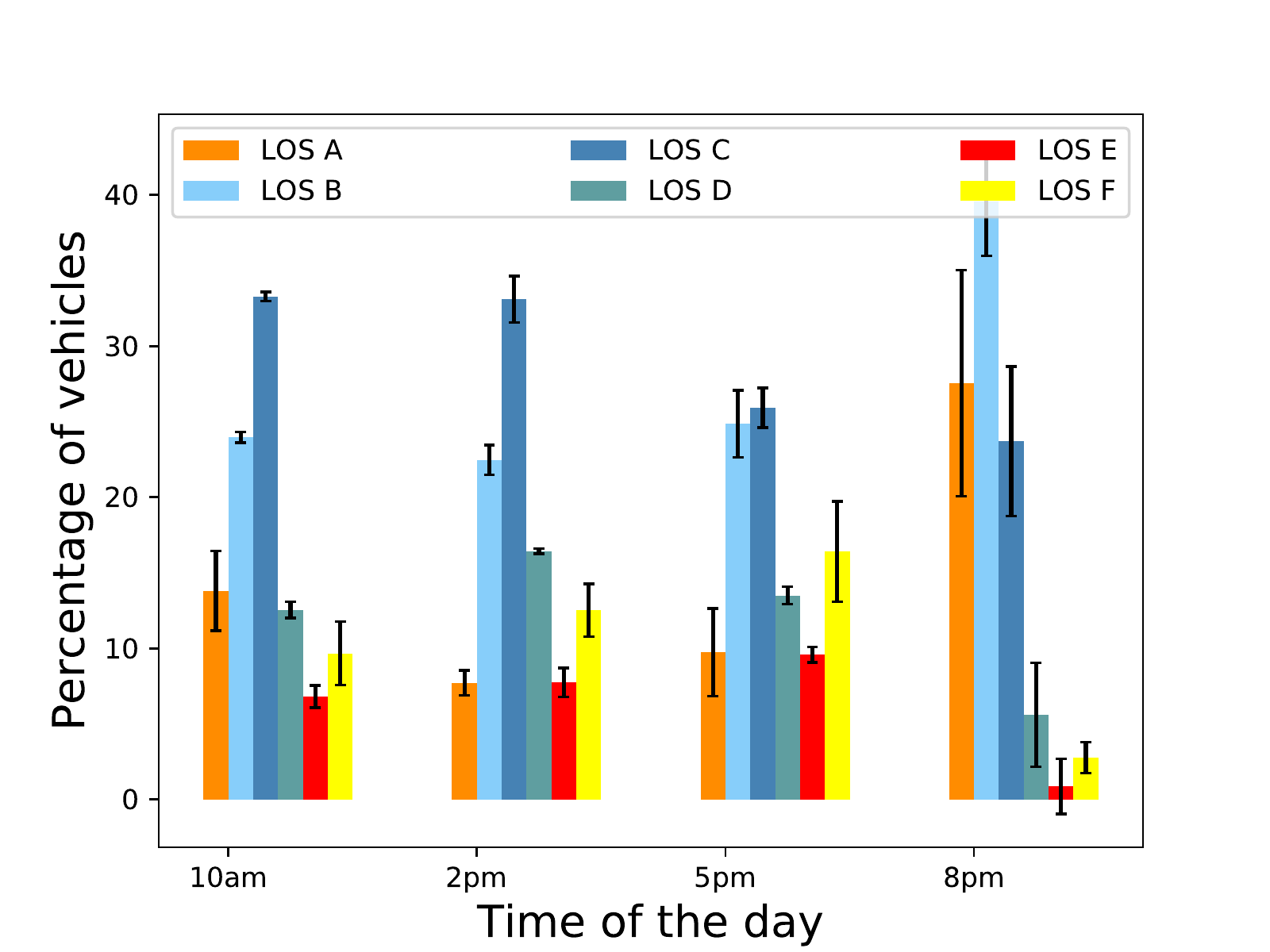}
   \end{center}
     \caption{The level of service (LOS) parameter for the I5-N freeway to understand the traffic conditions at different times of the day. The y axis shows the percentage of vehicles in each grade of LOS. LOS A depicts virtually free flow to LOS F which represents breakdown condition}
    \label{fig:LOS_freeway}
    \setlength{\belowcaptionskip}{-10pt}

    
\end{minipage}
\end{figure}


\subsubsection{Density-based detection of slow moving traffic}
\label{sec:density}
\hl{Beyond the predictability of vehicle mobility patterns, it is crucial to consider the density of vehicles, or how close or spread apart are vehicles at a given time, for the purpose of service placement. We take another real-world dataset to analyze the density and the traffic conditions for a freeway. Calculating density requires the number of vehicles in a given lane or calculated over a mile, averaged over a period of time. Density can also be calculated using occupancy of vehicles which is calculated as the percentage of time in which vehicles are passing over detectors. We take the data from the California Department of Transportation (Caltrans) Performance Measurement System (PeMS)\footnote{http://pems.dot.ca.gov/, available: 22/10/20.} that provides access to real-time and historical performance data for the California freeway network. This dataset and platform is a rich resource for many traffic parameters like occupancy, speed, vehicle miles travelled (VMT), etc. at many different granularities of time and space (over a detector/lane/route).}



\begin{table}[tb]
\caption{Relationship between density, volume and the level of service (LOS) on a freeway }
\label{tab:my-table}
\centering
\resizebox{\columnwidth}{!}{%
\begin{tabular}{llll}
\hline
\emph{LOS} &
  \emph{\begin{tabular}[c]{@{}l@{}}Density\\ (veh/mi/lane)\end{tabular}} &
  \emph{Volume/Capacity} &
  \emph{Traffic Flow Description} \\ \hline
A &
  0 \textless{}= x = 11 &
  x \textless{}= 0.3 &

  \begin{tabular}[c]{@{}l@{}}Virtually free flow; completely \\ unimpeded\end{tabular} \\ 
B &
  11 \textless{}= x \textless 18 &
  0.3 \textless x \textless{}= 0.5 &
  \begin{tabular}[c]{@{}l@{}}Stable flow with slight\\ delays;reasonably unimpeded\end{tabular} \\ 
C &
  18 \textless{}= x \textless 26 &
  0.5 \textless x \textless{}= 0.71 &
  \begin{tabular}[c]{@{}l@{}}Stable flow with delays; less\\ freedom to maneuver\end{tabular} \\ 
D &
  26 \textless{}= x \textless 35 &
  0.71 \textless x \textless{}= 0.89 &
  High density, but stable flow \\ 
E &
  35 \textless{}= x \textless 45 &
  0.89 \textless x \textless{}= 1.0 &
  \begin{tabular}[c]{@{}l@{}}Operating conditions at or near \\ capacity; unstable flow\end{tabular} \\ 
F &
  45 \textless{}= x &
  1.0 \textless{}= x &
  \begin{tabular}[c]{@{}l@{}}Forced flow, breakdown \\ conditions\end{tabular} \\ \hline
\end{tabular}
}
\end{table}

\hl{Vehicle-to-vehicle (V2V) communication is crucial for our distributed service placement and successful service completion. Reliability of V2V communication amongst vehicles in a cluster depends largely on traffic conditions. Conditions are indicated by vehicle density, which varies if vehicles are in a state of free flow, have delays, operating at full capacity, or are in breakdown condition. Caltrans PeMS provides a parameter called the Level of Service (LOS), which uses vehicle density to analyze the quality of service or the condition of the traffic along a freeway. The freeway LOS is a way to classify the traffic condition into a grading system ranging from A to F. The LOS is also measured from the volume-to-capacity (v/c) ratio, but PeMS estimates the LOS using density only. The relationship between LOS and density is defined in the Highway Capacity Manual (HCM2010) and is summarized in Table \mbox{\ref{tab:my-table}}. }

\hl{In Fig.\mbox{\ref{fig:LOS_freeway}} the percentage of vehicles in each LOS profile is plotted for different times of the day for the I5-N freeway for the first week in August 2019. As can be noted, there is a significant percentage of vehicles in LOS E which depicts unstable flow due to near full density of traffic, and LOS F which depicts flow breakdown conditions. The percentage of vehicles in LOS E and F constantly increases from 10 am to 5 pm. Almost 20\% of vehicles are in LOS F at 5 pm, which represents breakdown or traffic jam conditions. The percentage of vehicles increases significantly in LOS A and B at 8 pm, which depicts free flow and a stable flow of traffic. This analysis gives an estimate of available vehicle density at different times of the day. This reduces the uncertainty in resource availability. The data is a rich resource to detect busiest intersections and bottlenecks where services can be initiated on slow-moving or stopped vehicles.}

\subsection{Aggregate Communications Capacity Estimation}

\begin{table}[tb]
\caption{Aggregate capacity estimation for the I5-N freeway for Lane 1 based on the vehicular flow}
\label{tab:aggcap}
\resizebox{\columnwidth}{!}{%
\begin{tabular}{cccc}
\hline
\emph{\begin{tabular}[c]{@{}c@{}}Time of the\\ day\end{tabular}} & \emph{\begin{tabular}[c]{@{}c@{}}Vehicular flow\\ / lane1\end{tabular}} & \emph{\begin{tabular}[c]{@{}c@{}}LOS Profile\\ / lane1\end{tabular}} & \emph{\begin{tabular}[c]{@{}c@{}}Nominal \\ aggregate\\ capacity\\ / lane1\end{tabular}} \\ \hline
10 am & 50 & LOS F & 25 Mb/s \\ 
2 pm & 72 & LOS F & 36 Mb/s \\ 
5 pm & 61 & LOS F & 30.5 MB/s \\
8 pm & 63 & LOS F & 31.5 Mb/s \\
11 pm & 31 & LOS D & 15.75 Mb/s \\ \hline
\end{tabular}
}
\end{table}

Due to the novelty of using moving vehicles as infrastructure, we estimate the communication capacity of a vehicular network. 
\hl{Estimating the capacity of a vehicular network is a challenging problem to solve as it depends on several factors including the average number of simultaneous transmissions, link capacities, the density of vehicles, mobility in the network, the distance between vehicles, and the transmission range of the vehicles. Our previous analysis shows that the problem of less vehicular density causing a delay in communication is not prevalent in urban centers, and even freeway traffic flow in some cases. We also demonstrated that most traffic flow prediction can be done effectively. The  estimation of the capacity of the vehicular network has been done in great detail via customized theoretical studies \cite{916631,123,8039511}. We calculate the effective capacity of the vehicular network obtained using a cooperative scheme from Chen et al.\cite{8039511}. } 





\hl{\textbf{Theoretical Capacity}: We consider the closed-form expression of effective available capacity specified by Chen et al.\cite{8039511}, which uses a cooperative scheme to derive the communication capacity for a vehicular network. The cooperative strategy uses both V2V and Vehicle-to-Infrastructure (V2I) communication to increase the capacity of vehicular networks. They built an analytical framework to model the data dissemination process and derive a closed form expression of the achievable capacity, given as: }
    
        \begin{equation}
       \footnotesize{
              \begin{split}
            theoretical\_cap = \frac{L}{d}\min \{ W_I(1 - \exp^{-2\rho r_I}),\\ 
             W_I(1 - \exp^{-p \rho 2 r_I}) \\  + \frac{W_V.c_2(d - 2r_I)}{c_2.R_C + p -p\exp^{-2pr_o}} + \exp^{-p\rho2r_o} \}
             \end{split}
             }
        \end{equation} where $c_2 = (1-p)p \rho (1-\exp^{-\rho2r_o})$.
       \hl{In this expression $L$ is the length of the highway segment, d is the distance between RSUs, $W_I$ and $W_V$ are the data rate for V2I and V2V communication respectively, $\rho$ is the density of vehicles per meter, $p$ is the proportion of vehicles with download requests in the range [0,1], $r_I$ is the range of infrastructure points and $r_o$ is the radio range of vehicles. $R_C$ is the sensing range for the medium access control protocol. We calculate the available capacity for this case, taking the value for $L$ as 100 km, $d$ as 5, 10 or 15 km, $W_I$ as 20 Mb/s, $W_V$ as 2 MB/s, $\rho$ as 0.03, 0.04, or 0.05. We take the radio ranges as typical values for Dedicated Short-Range Communication (DSRC) such that $r_I$ is 400 m and $r_o$ is 200 m. The value of $R_C$ is taken as 300-400 m. For these values, the effective available capacity lies in the range of 5-20 Mb/s with different proportions of vehicles participating in the scheme. The density of vehicles, the use of cooperation schemes and the number of participating vehicles have a direct impact on this effective available capacity.}

 The potential computation capacity of a vehicle cluster is dependent on how dense the cluster is, in terms of the number of vehicles that are optimal for placement of a particular service request. The computation capacity is also based on how slow the vehicle cluster is, which can be predicted by the occupancy of a road segment, calculated as how much time vehicles take to pass over a detector. This time can also be derived as the sojourn time of vehicles with the RSU. According to the study conducted by Xiao et al.\cite{XIAO2019742}, predicted computation capacity is higher than 650 Gflops with a probability of 60\% when the range of vehicle clusters is set to be 5m, and throughout the day the computation capacity is above this value. When the range is 10m, the predicted capacity is 1800 Gflops. With the increasing number of smart vehicles, the number of sensors, video cameras, and computation capacity should increase significantly in the next decade. This means that the infrastructure will exist to collect data, process it on the resource pool of a vehicular cluster and send it to the cloud for further processing. However, this infrastructure cannot be exploited unless services can be placed on it in such a way that the overall service objectives are met.

 \subsubsection{Predicting aggregate computation capacity in a vehicular network}

The potential computation capacity of a vehicle cluster is dependent on how dense the cluster is, in terms of the number of vehicles that are optimal for placement of a particular service request. The computation capacity is also based on how slow the vehicle cluster is, which can be predicted by the occupancy of a road segment, calculated as how much time vehicles take to pass over a detector. This time can also be derived as the sojourn time of vehicles with the RSU. According to the study conducted by Xiao et al.~\mbox{\cite{XIAO2019742}}, predicted computation capacity is higher than 650 Gflops with a probability of 60\% when the range of vehicle clusters is set to be 5m, and throughout the day the computation capacity is above this value. When the range is 10m, the predicted capacity is 1800 Gflops. With the increasing number of smart vehicles, the number of sensors, video cameras, and computation capacity should increase significantly in the next decade; one estimate is for 150 million connected cars by  the end of 2020 \cite{8641431}. This means that the infrastructure will exist to collect data, process it on the resource pool of a vehicular cluster and send it to the cloud for further processing. However, this infrastructure cannot be exploited unless services can be placed on it in such a way that the overall service objectives are met.

\subsection{Microscopic and Macroscopic traffic trajectory data}

The utilization of the predictable vehicle trajectory and available processing capacity requires knowledge of both the microscopic behavior of individual vehicles as well as the overall macroscopic traffic flow dynamics. 

The microscopic trajectory of individual vehicles cannot be used because of privacy concerns over the use by third parties of user trajectory data. Therefore, to reduce privacy concerns, microscopic data can be considered at and between specific intersections, so there is no need to know the trajectory for the entire journey of a vehicle. This method can be predicted as the joint probability of vehicles starting at a road segment, say $RS_{j}$, and ending at the road segment say $RS_{k}$, expressed as:
\begin{equation}
\begin{split}
   P(src = RS_{j}, dest = RS_{k}) \\
   = P(src = RS_{j}).P(dest = RS_{k})   
\end{split}
\end{equation}




\noindent 
The macroscopic traffic data includes flow level variables like traffic flow rate, traffic density, and average velocity of the traffic stream. This data is easier to collect, using the vehicle counter and cameras commonly installed in cities for traffic management purposes. Macroscopic models also include deriving the relationship between traffic speed, flow rate, and density to estimate slow-moving vehicle traffic to initiate vehicle clusters. Most traffic estimation studies utilize both microscopic and macroscopic data to estimate vehicle trajectories. We have calibrated the microscopic car-following model, using the macroscopic vehicle flow data from the Dublin intersection based on the flow model in  \S\mbox{\ref{sec:flowmodel}}. For simulations, we extract the Dublin intersection road network using Open Street Map (OSM) and calibrate the simulation using the real-world Dublin traffic dataset. We generate the calibrated traffic in the Simulation for Urban Mobility (SUMO) simulator.

\subsection{Vehicular Fog Marketplace}

The problem of deploying edge servers and utilizing the traffic density in urban centers can be resolved by introducing a Vehicular Fog marketplace, where vehicles can temporarily lease some of their video capturing, sensing, computing, and networking capabilities. This marketplace would include \emph{consumers} in the form of service providers looking for reliable vehicular resources to capture and process information. The computational resources they seek can be used for applications that go beyond the motivating use case (for this paper) of crowdsourcing. Such additional use cases include intensive machine learning applications that can be implemented in a distributed manner. The \emph{producers} in the form of participating vehicles offer to host applications that pay a fair price while leasing the least amount of resources. The service provider aims to process most of the information on the vehicle cluster and collect as much data as possible, subject to the limitations of the infrastructure made available, collectively, by the vehicles in the cluster.

This IoV marketplace has been explored in the context of implementing complex, distributed machine learning models \mbox{\cite{8845247}}. The approach has been compared to job completion time with third party cloud providers. A similar marketplace needs to be studied for a moving cluster, in terms of monetary cost and task satisfaction.

\noindent 

\section{System Model}

In this section we first describe the terminology of the system model; then we present the the network topology and the distributed service model. 

\begin{figure}[t!]\centering
\includegraphics[width=0.9\linewidth,height=8cm]{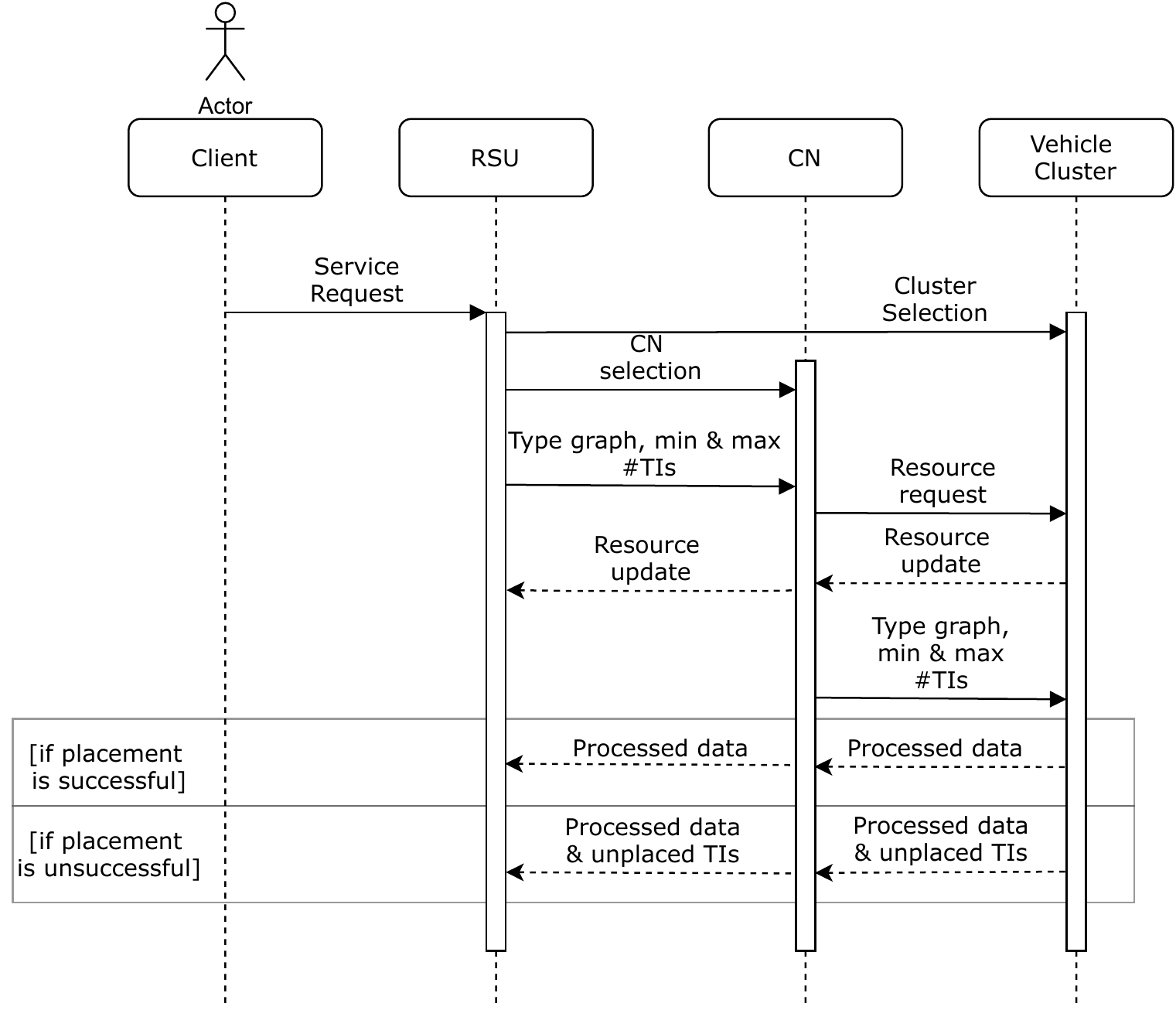}
\caption{System Model depicting the management between the RSU, CN and the vehicle cluster}
\label{fig:Management}
\end{figure}

\subsection{Terminology}

\begin{figure}[t!]\centering
\includegraphics[width=0.9\linewidth,height=5cm]{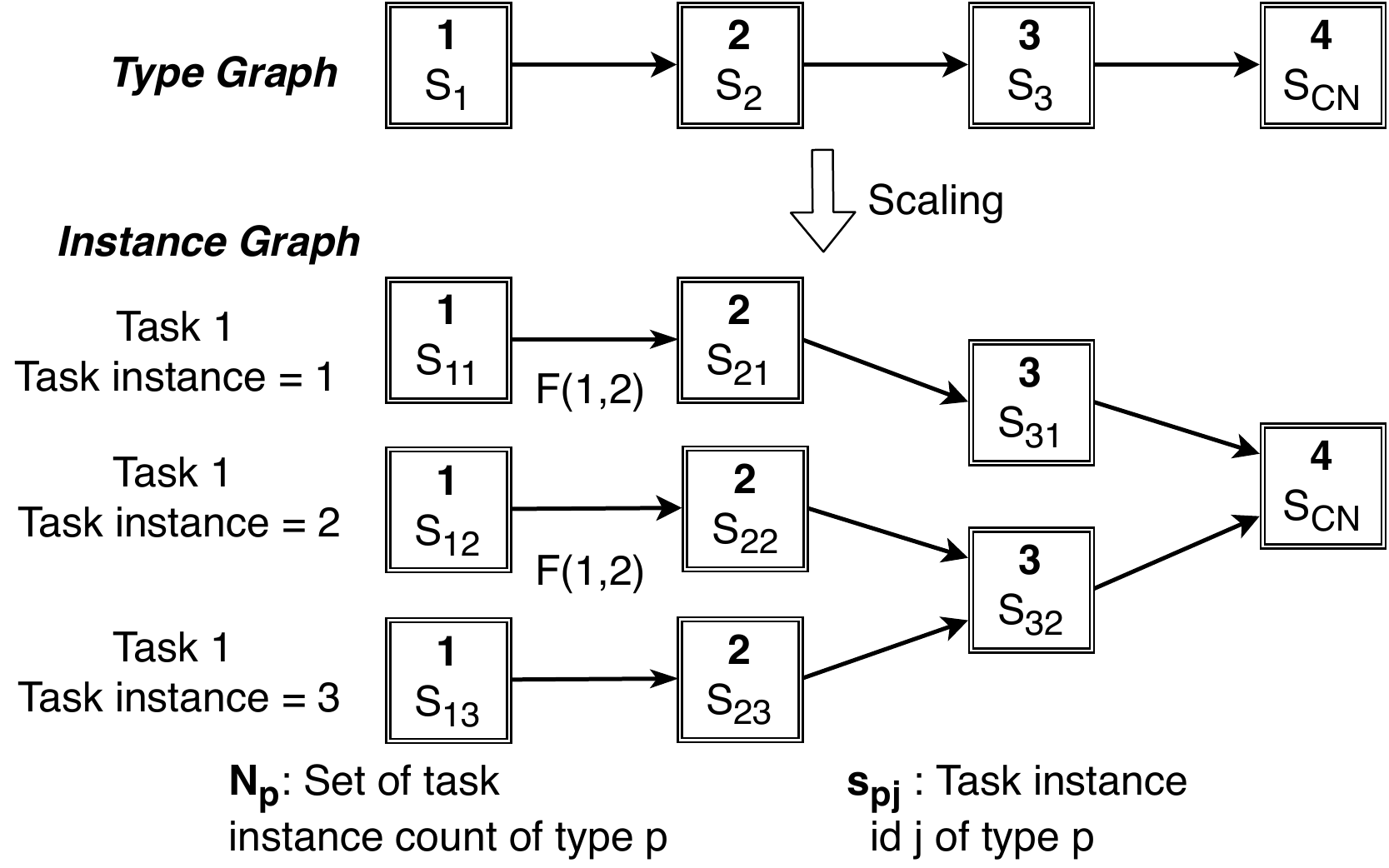}
\caption{Service model depicting tasks and their inter-dependencies. The Type graph is scaled to the Instance Graph based on the resource state of the vehicle cluster.}
\label{fig:ServiceModel1}
\end{figure}

\begin{itemize}
    \item \hl{\textbf{Vehicle Clusters}: We consider vehicle clusters as micro cloud-like entities \cite{8417759}, whose members (vehicles) provide resources used to execute tasks that form a distributed service.}
    \item \hl{\textbf{Control Node (CN)}: The CN is a vehicle in the cluster that acts as a gateway between the cluster and Roadside Units (RSUs); is elected based on its connectivity to the RSU and other cluster nodes (this election process is outside the scope of this paper). It collects  status information about the cluster, including available resources at nodes, link capacities and it also receives service placement requests from the RSU/client.}
    \item \hl{\textbf{Roadside Units (RSUs)}:  The vehicle nodes in a cluster are supported by resource-rich base stations (RSUs), which connect the cluster to the Internet. The management of services between the RSU, Control Node (CN), and the vehicle cluster is depicted in \mbox{Fig.~\ref{fig:Management}}. The RSU knows the system state of the cluster, which is communicated to it by the CN.}
    \item \hl{\textbf{Task}: Tasks are data collection or processing functions that can be scaled out as multiple task instances (TIs) to realise a distributed service. For example, a distributed service to realise pedestrian counting may in its specification request as many vehicle cameras as possible monitoring a given stretch of road. The TIs are the smallest unit that a task can be split into and that can be mapped to a vehicle node }
    \item \hl{\textbf{Service}: We consider distributed services with unidirectional, acyclic control, and data-flows. These services are specified as hierarchies of different task types, each with different functionality. Each task is typically deployed as several TIs, which can be dynamically and flexibly scaled (in terms of size per TI (\emph{up}) and number of TIs per task (\emph{out})) according to resource availability and stability of the vehicle cluster at a given instant. We assume a linear chain of data-dependent tasks  represented as a \emph{Type graph}, in \mbox{Fig.~\ref{fig:ServiceModel1}}. This Type graph is sent as an input to the service placement function. Based on the Type Graph, an \emph{Instance graph} is created, where each task of Type p (represented as $s_p$ in \mbox{Fig.~\ref{fig:ServiceModel1}}) can have multiple TIs of Type p and count j (represented as $s_{pj}$). Other works that leverage parked vehicles (PVs) also deploy similar service models, where a task with a large workload is split into several sub-tasks and assigned to multiple PVs for cooperative execution \cite{8939405}.}
    \item \hl{\textbf{Service Placement}: The process of placing the scaled Instance graph (in \mbox{Fig.~\ref{fig:ServiceModel1})} on a vehicle cluster, is called the service placement problem. }
\end{itemize}

Our approach is to first find an optimal \hl{Instance graph}, considering both service and infrastructure constraints, as this decision cannot be taken independently of the infrastructure state. This is optimized based on minimizing the total number of hops in the path between each \emph{Type 1} TI and the CN. This step reduces the bandwidth usage and selects a dense service spread, which also reduces delay in service execution. We then map the optimized Instance graph onto the physical vehicle nodes. \hl{We jointly consider both TI mapping as well as the route/flow mapping\mbox{~\cite{8082507}}, between the placed TIs. We also take into account the predictable mobility pattern of these vehicles.}

\subsection{Network Topology}

We assume that $\mathbb{I}$ nodes participate in the formation of the vehicle cluster. We represent the cluster as a directed, connected graph, $G = (V,E)$. The node $i\in V$ represents the vehicle nodes, each with $K$ resource types, where $k \in \{1...K\}$ and $i \in \{1...\mathbb{I}\}$ denote resource type $k$ on node $i$. The processing capacity of each vehicle node $i$ in respect of resource type $k$ is represented as $C_{k}(i)$. The directed edge, $(i_1,i_2) \in E$, of the graph represents the link between any two vehicle nodes $i_1$ and $i_2$. The link capacity limit is depicted as $\{B(i_1,i_2)\}$ Kb/s between any two nodes $i_1$ and $i_2$. If there is no direct connectivity, due to excessive range, line of sight difficulties and/or incompatible protocols then $B(i_1,i_2) \equiv 0$.

The mobility in the network is represented by the cluster cohesion probability (CCP) of each vehicle node ($P_{(t_1,t_2)}(i_1)$), which represents the probability of a vehicle to be in a certain segment of the road, in a particular period of time $[t_1, t_2]$. We also consider the joint probability ($P_{(t_1,t_2)}(i_1,i_2)$) of two nodes $i_1$ and $i_2$ to stay together on a given road segment over the time interval $[t_1, t_2]$, due to the inherent data dependency between two interacting \hl{task} nodes, as specified in the service model. This makes it important to consider the combined probability of two nodes with data-dependent \hl{TIs} to stay together until the completion of both \hl{TI} tasks. We assume that this information regarding the mobility pattern, in terms of CCP of the nodes, is available at each road intersection.

\subsection{Service Model: Task and Task Instances}

The service model is composed of \hl{tasks}, denoted as $s_{p}$, each with different functionality, to be deployed on different nodes of the cluster. The functions include video streaming, data compression/processing as well as application control, for the flexible management of infrastructure links and nodes. Each \hl{task} can have any number of \hl{TIs}, represented as $s_{pj}$, to be mapped in an optimal configuration onto vehicle nodes. The number of \hl{TIs} for \hl{a task} $s_{p}$ is represented as $\mathbb{N}_{s_{p}}$. Each \hl{TI} $s_{pj}$ requires a minimum demanded amount of $D_{pjk}$ units of each resource of type $k$. Furthermore, the flow \emph{demand} between \hl{task} $s_{p_1}$ and \hl{task} $s_{p_2}$ is provided as $F(s_{p_1},s_{p_2})$. Note that such flows might be point-to-point (between adjacent nodes) or might need to be routed via other nodes according to flow tables maintained by the CN. Both the per resource type $k$ demand for \hl{TI} $s_{pj}$, labeled by $\{D_{pj,k}\}$, and the inter-\hl{task} demand $\{F(s_{p_1},s_{p_2}), s_{p_1}\neq s_{p_2}\}$ need to be specified as input to the model.


Each \hl{TI} can support a maximum flow rate, which is derived from the processing requirement of the incoming flow, given as $C(F(s_{p_1j},s_{p_2j}))$, i.e., the processing requirement for flow from \hl{TI} $s_{p_1j}$ to $s_{p_2j}$. We check that the target \hl{TI} has enough processing capacity to process an incoming flow and also ensure that this leaving flow, after being streamed or processed at an \hl{TI}, is directed to a single corresponding \hl{TI}. Recall that each processing \hl{TI} can have multiple incoming flows. We follow this rule to promote the collocation of processing nodes, whenever vehicle nodes have available resource capacity. This promotes a balanced service placement rather than over-provisioning the available infrastructure. 

We aim to minimize the cost of service execution, by favoring nodes with a higher probability of staying with the vehicle cluster and promoting a ``narrower'' service placement (just enough nodes for reliability in the presence of node mobility) to reduce resource bandwidth usage. The model can be used to optimize other resources like the increasing number of accepted requests on vehicle clusters, the number of nodes used or other performance metrics like latency or service bandwidth demand, based on the requirements of the application.

\section{Constraints and Problem Formulation}

\hl{The placement problem first scales the service type graph to an instance graph and then \emph{maps} the service onto the vehicle cluster. }
\subsection{Infrastructure Constraints}
\subsubsection{\textbf{Node Resource Constraints}}

Each resource type of a vehicle node is denoted as $k$, where $k=1$ is CPU capacity, $k=2$ is Memory capacity and $k=3$ is sensing resource capacity. The minimum resource \emph{requirement} (of type $k$) to host \hl{TI} $s_{pj}$ on node $i$ is given as $D_{pj,k}$. This constraint \mbox{~\ref{eq:NodeResourceConstraint}} checks if the \hl{TI} is mapped to node $i$ which is denoted by the binary mapping variable $M(p,j,i)$. The resource required by the placed \hl{TI} must not exceed the availability of resource type $k$ on the selected node. Since we are interested in using only spare resources for placing services on vehicular clusters, a minimum set of system resources must also be reserved for the operations required for the vehicles. Thus, the net available capacity for hosting collaborative services at every node $i$ is represented as $C_{k}(i)$, which is the \emph{capacity} of the node for such additional services. The resource constraint is formally presented as: 
.
\begin{equation}
\small{
\begin{split}
\forall{i \in \{1,\ldots,\mathbb{I}\}, k \in \{1,\ldots,\mathbb{K}\}},\sum_{\forall p,j} M(p,j,i).D_{pj,k} \leq C_{k}(i)
\end{split}
}
\label{eq:NodeResourceConstraint}
\end{equation}
where the decision variable $M(p,j,i) \in \{0,1\}$ is set to 1 to indicate that \hl{TI} $s_{pj}$ is placed on node $i$ or 0 otherwise. 
The use of this indicator variable ensures that \hl{TI} $s_{pj}$ requires resources from node $i$, \emph{if and only if} it is placed on that node.

\subsubsection{\textbf{Bandwidth Constraint}} The bandwidth requirement between two \hl{task} $s_{p_1}$ and $s_{p_2}$, where the latter requires data from the former, is represented by $F(s_{p_1}, s_{p_2})$ Kb/s. We consider only one-directional traffic, from \hl{task} $s_{p_1}$ \emph{to} $s_{p_2}$. However, the model can easily be extended to consider duplex communication needs by adding extra constraints of the form~\ref{eq:BandwidthConstraints}. 
\begin{equation}
\footnotesize{
\begin{split}
\forall{i_1 \in \{1,\ldots,\mathbb{I}\}; i_2 \in \{1,\ldots,\mathbb{I}\}; i_1 \neq i_2}\\
\sum_{\forall{p_1,j_1; p_2,j_2; p_1 \neq p_2}} M(p_1,j_1,i_1)F(s_{p_1},s_{p_2})M(p_2,j_2,i_2)\leq  B(i_1,i_2)
\end{split}
}
\label{eq:BandwidthConstraints}
\end{equation}
where $i_1 \neq i_2$ and $B(i_1,i_2) \neq 0$. Constraint \mbox{~\ref{eq:BandwidthConstraints}} ensures that, for each node pair labeled by $i_1$ and $i_2$, the total bandwidth requirement, for all \hl{TI} pairs $s_{p_{1}j}$ and $s_{p_{2}j}$ placed on nodes $i_1$ and $i_2$ respectively, is $F(s_{p_1},s_{p_2})$, which does not exceed the bandwidth limit $B(i_1,i_2)$ between the two nodes. 

In our model, two \hl{tasks} that are mapped to two different vehicle nodes $i_1$ and $i_2$, where $i_1 \neq i_2$ might not be linked directly to each other, but are connected over multiple hops. In the following bandwidth constraint, we consider the resource capacity of each link over the full path between \hl{tasks} $s_{p_1}$ and $s_{p_2}$. We consider another binary valued mapping variable $m(p_1,p_2,j,i)$ which takes the value 1 for each node $i$ that is mapped to forward the flow between \hl{TIs} of type $s_{p_1j}$ and $s_{p_2j}$  and is part of the path between the two data dependent \hl{TIs}. Thus, nodes can act as both processing nodes or forwarding nodes. The constraint \mbox{~\ref{eq:BandwidthConstraints_multihop}} ensures that the bandwidth used for forwarding the flow between any connected pair of forwarding nodes should be less than the available bandwidth capacity between those two nodes. This constraint is formally presented as:
\begin{equation}
\small{
\begin{split}
\forall{i_1 \in \{1,\ldots,\mathbb{I'}\}; i_2 \in \{1,\ldots,\mathbb{I'}\}; i_1 \neq i_2}\\
\sum_{\forall{p_1,j_1; p_2,j_2;p_1 \neq p_2}} m(p_1,p_2,j_1,i_1)F(s_{p_1},s_{p_2})m(p_1,p_2,j_1,i_2)\\
\leq  B(i_1,i_2)
\end{split}
}
\label{eq:BandwidthConstraints_multihop}
\end{equation}

where $i_1$ and $i_2$ belong to $\mathbb{I'}_{(p_1j)(p_2j)}$, which is the set of all nodes on the path between \hl{TIs} $s_{p_1j}$ and $s_{p_2j}$. \\ 






\subsection{\textbf{Service Model Constraints}}

We now formulate the constraints for placing distributed \hl{TIs} and the corresponding service data flow between these \hl{TIs}. We ensure that the data flow is processed before reaching the CN and the order of \hl{TIs} is maintained according to the service chain or service description.

\subsubsection{\textbf{Flow Rate Constraint}} As we propose a distributed service model, it is crucial to ensure that the \hl{TIs} have enough processing capacity for the incoming flow. The constraint \mbox{~\ref{eq:FlowRateConstraints}} ensures that the flow rate entering a \hl{TI} should not exceed the processing capacity of that \hl{TI}. The processing capacity of \hl{TI} $s_{p_2j}$ is represented as $C(F(s_{p_1j},s_{p_2j}))$, which is the function of incoming flow from $s_{p_1j}$ to $s_{p_2j}$. This constraint is given as:
\begin{equation}
\small{
\begin{split}
\forall{i_1 \in \{1,\ldots,\mathbb{I}\}; i_2 \in \{1,\ldots,\mathbb{I}\}; i_1 \neq i_2}\\
\sum_{\forall{p_1,j_1; p_2,j_2; p_1 \neq p_2}} m(p_1,p_2,j_1,i_1)C(F'(s_{p_{1}j},s_{p_{2}j})) \leq C(s_{p_{2}j})
\end{split}
}
\label{eq:FlowRateConstraints}
\end{equation}
\noindent
where $C(s_{p_{2}j})$ represents the processing capacity of \hl{TI} $s_{p_2j}$. This \hl{TI} is placed on the node that receives the incoming flow to be processed, from \hl{TI} $s_{p_1j}$. Here $F'(s_{p_1j},s_{p_2,j})$ represents the flow that has been processed at \hl{TI} $s_{p_1j}$ or is forwarded from $s_{p_1j}$ specifically for processing (not forwarding). 

\subsubsection{\textbf{Flow Conservation Constraint}} Constraint \mbox{~\ref{eq:FlowConservationConstraints}} ensures that the incoming to outgoing flow rate ratio, at a node, is governed by the data processing factor of the \hl{TI}. $\alpha_{pj}$ represents the data reduction/processing factor for a \hl{task} with \emph{Type p} resource. The constraint is presented as: 
\begin{equation}
\small{
\begin{split}
\forall{i_1 \in \{1,\ldots,\mathbb{I}\}; i_2 \in \{1,\ldots,\mathbb{I}\}; i_1 \neq i_2}\\
\sum_{\forall{p_1,j_1; p_2,j_2; p_1 \neq p_2}} F(s_{p_1j},s_{p_2j})\alpha_{p_2j} \leq F'(s_{p_1j},s_{p_2j})\\
\end{split}
}
\label{eq:FlowConservationConstraints}
\end{equation}
where $0 \leq \alpha_{pj_1} \leq 1$ and $F(s_{p_1j},s_{p_2j})$ represents the incoming flow to be processed at \hl{TI} $s_{p_2j}$. $F'(s_{p_1j},s_{p_2j})$ represents the outgoing flow, that has been processed at the \hl{TI} $s_{p_2j}$. This constraint ensures that all the necessary pre-processing is performed on the flow, at each \hl{TI} before the flow reaches the CN. 
Since nodes in our model can be forwarding nodes, or processing nodes
, or have both a processing and forwarding role, the data processing factor can lie in the range from [0,1].

\subsubsection{\textbf{Task Order Constraints}} Constraint \mbox{~\ref{eq:ServiceComponentOrderConstraints}} ensure that the flow traverses the \hl{task instance} graph in the order specified by the service model, we require that once the flow is processed at one node, it is directed to the ``next'' node with at least one ``subsequent'' \hl{TI} (according to the Type Graph), i.e.,
\begin{equation}
\small{
\begin{split}
\forall{i_1 \in \{1,\ldots,\mathbb{I}\}; i_2 \in \{1,\ldots,\mathbb{I}\}; i_1 \neq i_2}\\
\sum_{\forall{p,j; p+1,j_2; p \neq p+1}} 
M(p,j,i_1)F'(s_{p_1j},s_{p_2j}) \geq M(p+1,j,i_2)
\end{split}
}
\label{eq:ServiceComponentOrderConstraints}
\end{equation}
where the decision variable $M(p,j,i_1)$ represents the \hl{TI} mapped to node $i_1$, $F'(s_{p_1j},s_{p_2j})$ is the flow processed at the node $i_1$. The right hand side of the equation employs mapping instance $M(p+1,j,i_2)$ to show that a subsequent \hl{TI} of type $s_{(p+1)j}$ is mapped on the node $i_2$, which has enough resource capacity. 

As forwarding nodes are introduced in constraint \mbox{~\ref{eq:BandwidthConstraints}} to facilitate these multi-hop flows, it is crucial to preserve the order of \hl{tasks} at the service level. To ensure that the flow is directed towards a subsequent \hl{TI}, in case there is no direct path between two placed \hl{TIs}, we also ensure that the forwarding node is on the path joining nodes $(i_1,i_2)$ with \hl{TIs} $s_{pj}$ and $s_{(p+1)j}$ mapped on them. This is represented as constraint \mbox{~\ref{eq:ServiceComponnetOrderConstraints2}}:
\begin{equation}
\small{
\begin{split}
\forall{i_1 \in \{1,\ldots,\mathbb{I}\}; i_2 \in \{1,\ldots,\mathbb{I'}\}; i_1 \neq i_2}\\
\sum_{\forall{p,j; p+1,j_2; p \neq p+1}} 
m(p,p+1,j,i_1)F'(s_{p_1j},s_{p_2j}) \\
\geq m(p,p+1,j,i_2)
\end{split}
}
\label{eq:ServiceComponnetOrderConstraints2}
\end{equation}
where $m(p,p+1,j,i_1)$ and $m(p,p+1,j,i_2)$ are mapping variables with 0 or 1 value. Here $m(p,p+1,j,i_1)$ represents node $i_1$ as a forwarding node for processed data flow $F'(s_{p_1j},s_{p_2j})$, between \hl{TIs} $s_{pj}$ and $s_{(p+1)j}$, and $m(p,p+1,j,i_2)$ represents the next forwarding node for the same flow. 

\subsection{Cluster cohesion probability}
\label{sec:CCP}
In order to use the mobility of slow moving vehicles in our favor, it is crucial to incorporate mobility awareness in the infrastructure model. There are many ways to predict the mobility patterns of a group of vehicles. Here we consider all nodes that have a higher probability to chose a similar road segment ($S_i$), based on their historical mobility patterns, to be candidates for the cluster. We assume that each RSU maintains a table of known vehicle nodes, with their probability of taking a particular road segment (say $S_1$) at the next intersection. Vehicles that do not have entry in the table, but are willing to offer their resources, can be added to the table. However, they would be assigned the average road exit probabilities of known vehicles, with a low confidence score. As the history of a given vehicle builds up with time, its road exit probabilities are updated and their confidence score increases.

\begin{figure}[tb]
\begin{minipage}[t]{\linewidth}
\begin{center}
    \includegraphics[width=0.9\linewidth,height=5cm,keepaspectratio]{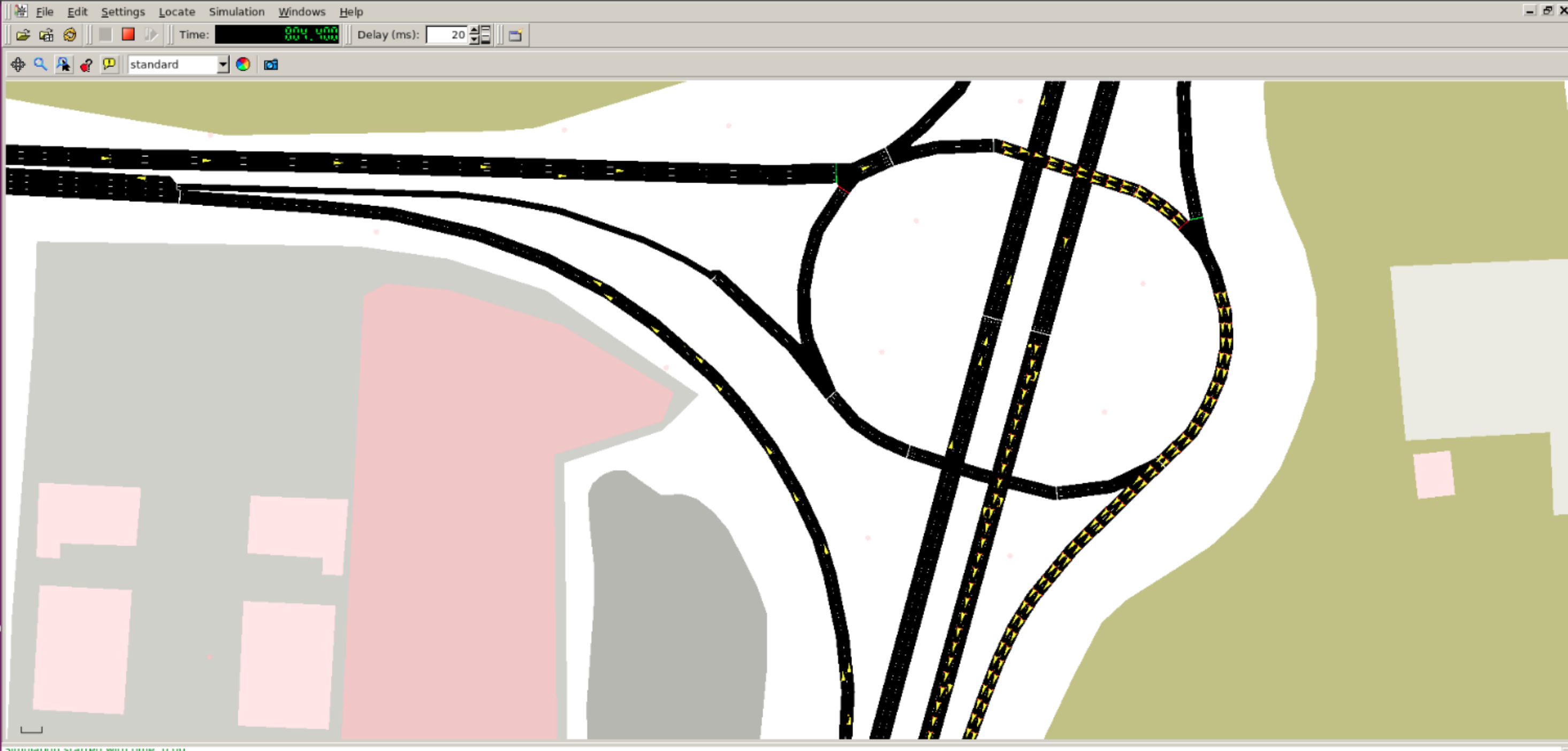}
    \end{center}
    \caption{The SUMO simulation for the intersection in Dublin, Ireland}
    \label{fig:sumo}
\end{minipage}
\end{figure}
 We have calibrated the microscopic car-following model, using the macroscopic vehicle flow data from the Dublin intersection based on the flow model in \S\mbox{\ref{sec:flowmodel}}. For simulations, we extract the Dublin intersection road network, as depicted in Fig. \mbox{~\ref{fig:sumo}}, using Open Street Map (OSM) and calibrate the simulation using the real-world Dublin traffic dataset. We generate the calibrated traffic in the Simulation for Urban Mobility (SUMO) simulator. 

\newcommand{\matindex}[1]{\mbox{\scriptsize#1}}
\[
  A=\begin{blockarray}{ccccc}
    \matindex{$S_1$} & \matindex{$S_2$} & \matindex{$S_3$} & \matindex{$S_4$} & \\
    \begin{block}{(cccc)c}
      P_{(t_1,t_2)}(car1) & \dots & \dots & \dots & \matindex{$car \ 1$}\\
      0.5 & 0.4 & 0.1 & 0 & \matindex{$car \ 2$} \\
      0.4 & 0.6 & 0 & 0.1 & \matindex{$car \ 3$} \\
      0 & 0 & 0.9 & 0.1 & \matindex{$car \ 4$}\\
          0.5 & 0.5 & 0 & 0 & \matindex{$car \ 5$}\\
            0.6 & 0.4 & 0.2 & 0 & \matindex{$car \ 6$}\\
    \end{block}
  \end{blockarray}
\]

The transition matrix stores the mobility behavior of every candidate vehicle, for a particular time period. This table can be updated over time to increase the accuracy of mobility awareness. Each RSU thus has many tables stored for different time stamps during the day. We model the mobility of vehicles as a Markov Model, where each road segment is a state. As mentioned in \cite{7015614}, the vehicle node that moves from one road segment to the other represents a transition in the Markov process. But instead of considering the detailed trajectory of a single vehicle, the matrix stores all possible probabilities for a vehicle to stay at the segment or take another road segment with a certain probability. Thus, every intersection in the service zone maintains the probability of a vehicle that follows Markov memory-less property, wherein the node transitions from state i to i+1 and is independent of state i-1. Based on the mobility patterns, different vehicle clusters can be formed for the service execution. In this paper, we only consider the nodes with a high probability of going from road segment A to C (in Fig.~\ref{fig:flowmodel}), as continuing to belong to the cluster. Therefore, the cluser cohesion probability (CCP) of a given vehicle node is the probability of that node going straight ahead at the next intersection. 


\subsection{Service Placement Cost}

To incorporate the mobility of hosting nodes, we scale the resource capacity of each node in the \hl{vehicle} cluster with a weighting factor, i.e., the probability of a node to stay with the cluster for the duration of service execution, i.e., from time $t_1$ to $t_2$, which is given as $P_{(t_1,t_2)}(i)$. This is because a node with enough resource capacity might not have a high probability of staying with the vehicle cluster, so this needs to be considered when placing \hl{TIs} on that node. Placing \hl{TIs} on such nodes can waste computation and bandwidth resources if the node leaves the cluster prematurely, and can also cause the service to fail. Thus, we scale the vehicle node capacity with its CCP, such that the \emph{higher} the CCP (probability of staying with the cluster), the \emph{lower} the costs of \hl{TI} execution on that node. The Node Cost is given as:
\begin{equation}
\footnotesize{
\begin{split}
  \mbox{NodeCost} (i_1) = \sum_{\forall{i_1, i_2; i_1 \neq i_2}}\\ (1 - P_{(t_1,t_2)}(i_1)).(D_{pj,k}/C_{k}(i)).M(p,j,i)
\end{split}
}
\label{eq:NodeCapacityCost}
\end{equation}
where $M(p,j,i)$ is the mapping function of \hl{TI} $s_{pj}$ to node $i$, with node resource capacity of $C_{k}(i)$. To add the costs, we consider the ratio of required node capacity $(D_{pj,k})$ with the available node capacity $(C_{i}(k))$. 

Similarly we scale the link capacity of any two nodes with data-dependent \hl{TIs}, with the joint probability of the two nodes to stay together for the duration of service execution ($t_1$ to $t_2$), given as $P_{(t_1,t_2)}(i_1,i_2)$. The total link cost for service execution is given as:
\begin{equation}
\footnotesize{
\begin{split}
  \mbox{LinkCost} (i_1,i_2) =
\sum_{\forall{i_1, i_2; i_1 \neq i_2}} (1 - P_{(t_1,t_2)}(i_1,i_2)).\\ (F(p_1,p_2)/B(i_1,i_2))\Big( m(p,p+1,j_1,i_1).m(p,p+1,j_2,i_2) \\
+ M(p,j_1,i_1).M(p,j_2,i_2)\Big)
\end{split}
}
\label{eq:LinkCost}
\end{equation}

where $m(p,p+1,j_1,i_1).m(p,p+1,j_2,i_2) \in \{0,1\}$ is an indicator that two nodes that form part of the path joining two \hl{TIs} of \hl{task} $s_{p_1j}$ and $s_{(p+1)j}$ type. Similarly $M(p_1,j,i_1).M(p_2,j,i_2)$ indicates that two nodes, one hosting \hl{TI} $s_{p_1j}$ at node $i_1$ and the other \hl{TI} $s_{p_2j}$ at node $i_2$, have a direct link between them. For adding up the link cost and the operating cost on each node, we use the ratio of required bandwidth resource $(F(s_{p_1j},s_{p_2j}))$ with the available bandwidth $(B(i_1,i_2))$ at each link that forms part of the service placement.



\subsection{Objective Function}
The problem is formulated as a bi-objective optimization. We hierarchically solve the optimization with the first objective as:

\subsubsection{\textbf{Adjacency \hl{TI} placement}}



When placing \hl{tasks} on nodes, it is more efficient to ensure that the placement plan takes account of both \hl{task} dependencies and of inter-node network distances. For example, if $s_{p_2}$ depends on $s_{p_1}$, it is advisable to ensure that each is placed either on the same node or on nodes that are one hop away from each other. However, this requirement could make it difficult to find a feasible placement. Hence, we seek to ensure that the network distance between any two selected nodes with data dependency is minimized for efficient service placement. The hop count between two placed \hl{TIs} is minimized when:
\begin{equation}
\footnotesize{
\begin{split}
\forall{i_1 \in \{1,\ldots,\mathbb{I'}_{(p_1j)(p_2j)}\}; i_2 \in \{1,\ldots,\mathbb{I'}_{(p_1j)(p_2j)}\}; i_1 \neq i_2}\\
H(i_1,i_2) = \sum_{\forall{p_1,p_2}} m(p_1,p_2,j,i), \min  \ H(i_1,i_2)), \\
\end{split}
}
\label{eq:AdjacencyConstraints}
\end{equation}
where $H(i_1,i_2)$ is the hop count between two nodes $i_1$ and $i_2$ for the flow $F(s_{p_1j},s_{p_2j})$ between \hl{tasks} $s_{p_1}$ and $s_{p_2}$. In the model, mapping variable $m(p_1, p_2, j, i)$ applies to a forwarding node $i$ along the path between \hl{TIs} $j$ of type $p_1$ and $p_2$, in cases where there is no direct link between the nodes hosting these \hl{TIs}. 

\subsubsection{\textbf{Total Cost of Service Placement}}
\begin{figure}[tb]
\begin{minipage}[t]{\linewidth}
\begin{center}
    \includegraphics[width=8cm,height=5cm]{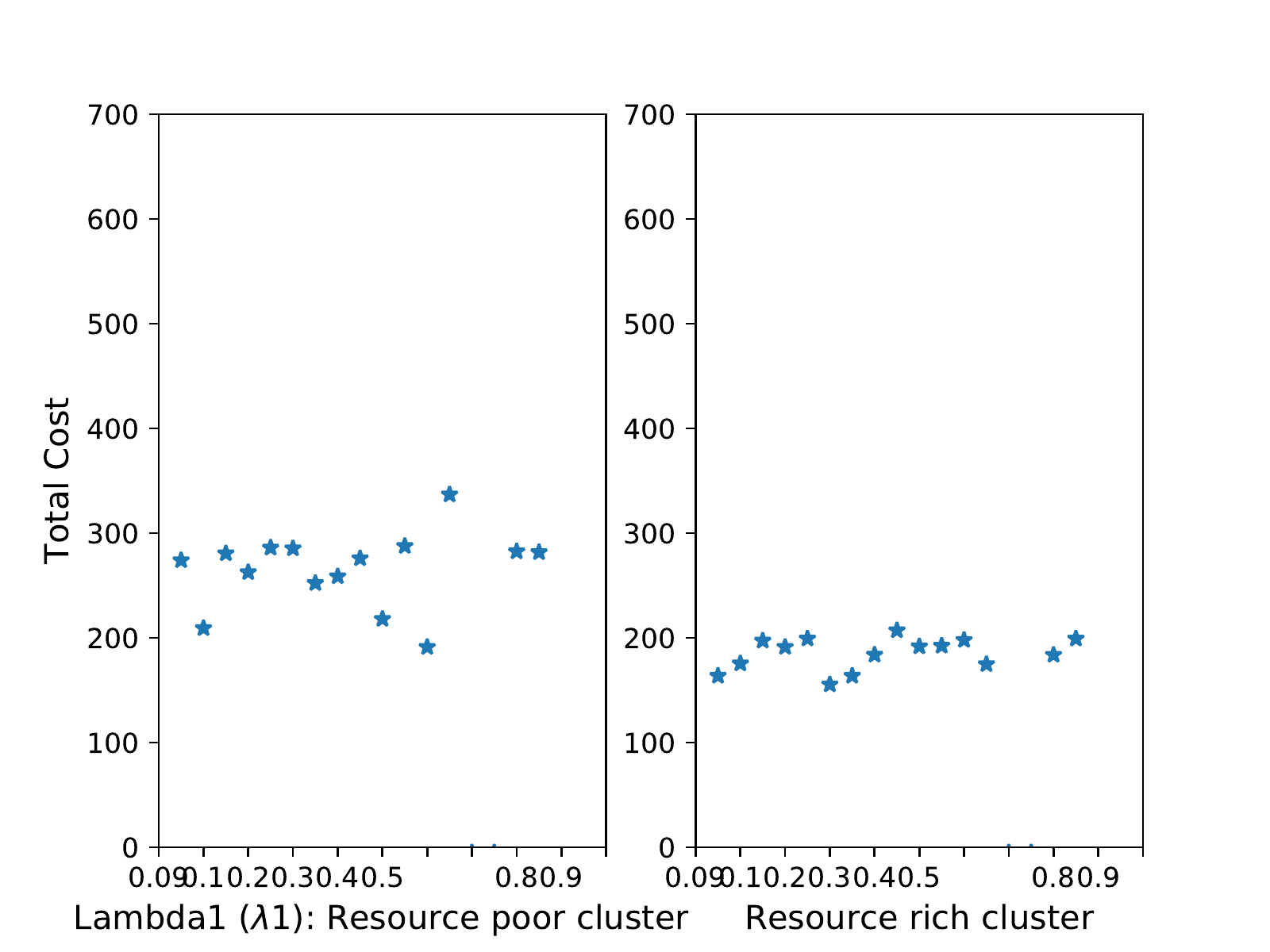}
    \end{center}
   \caption{Sensitivity analysis for resource-rich and resource poor cluster: we generate random values for $\lambda_1$ (on x-axis of the graph), and plot the total cost for the corresponding resource-poor and resource-rich clusters. In the event of low node capacity or low link capacity, the optimization takes care of the number of instances deployed, with the objective of minimizing total cost. Thus, as depicted in this figure, we get total cost values in a similar range for almost every weight. We chose both $\lambda_1$ and $\lambda_2 = 0.5$, as it gives lower cost in both cases and other values do not significantly affect cost.}
    \label{fig:weights}
\end{minipage}
\end{figure}

We then solve the model for the next objective function, which minimizes the Total Cost spent on service execution:
\begin{equation}
\footnotesize{
  \min\sum_{\forall{i_1, i_2; i_1 \neq i_2}} \lambda_1 \mbox{LinkCost} (i_1,i_2) + \lambda_2 \mbox{NodeCost} (i_1) \\
}
\label{eq:ObjectiveFunction}
\end{equation}
\hl{where $\lambda_i$ are non-negative and sum to 1. When evaluating our model, we set $\lambda_1 = \lambda_2 = 0.5$, i.e., we give equal weight to node and link cost for simplicity. To come to this decision, we carry out a sensitivity analysis and generate random values for $\lambda_1$ ($\lambda_2$ = 1 - $\lambda_1$), and plot total cost for the same resource-poor and resource-rich clusters. In the event of low node capacity or low link capacity, the optimization takes care of the number of TIs deployed, with the objective of minimizing total cost. Thus, as depicted in \mbox{Fig.~\ref{fig:weights}}, we get total cost values in a similar range for almost every weight. We chose both $\lambda_1$ and $\lambda_2 = 0.5$, as it gives lower cost in both cases and other values do not significantly affect cost.} Also, in hierarchical optimization, the first objective function effectively gets a higher priority than the next. We give an explanation of this choice in \mbox{Section~\ref{sec:Results}}. Thus, minimizing the hop counts is the first priority of the optimization and then equal priorities are given to both node and link cost.

\subsection{Problem Complexity}

The placement of a linear application graph on tree graph, for data center topology has been widely studied in the literature. In \cite{7847322}, they aim to minimize the maximum weighted cost on each physical node and link, ensuring that no single element gets overloaded and becomes a point of failure. They prove that the placement of a tree application graph onto a tree physical graph for the objective function, with or without pre-specified junction node placement, is NP-hard.

We first scale a linear \hl{Type} graph to a tree graph (Instance Graph) and then map it to a general directed graph (vehicle cluster). The mobile-edge computing application placement problem does not consider the changes in resource state due to the movement of infrastructure. Thus, the initial placement of service chains on a vehicle cluster can be considered similar to the application mapping problem. The application or workload placement problem is a hard problem to solve, especially while considering the joint node and link mapping. We extended this problem from a traditional NFV service chaining problem, which has fixed source and destination nodes, with a fixed length of the service chain. In our case, due to the lack of knowledge of individual vehicle's mobility dynamics, and the need for a robust placement, we scale the linear chain to multiple data collection (Type 1) \hl{TIs}. In our model, if the data collection \hl{TI} is unable to find a processing placement instance on the vehicle cluster, it forwards the data to the RSU. Thus, the model includes both horizontal and vertical scaling of the service. We also place more than one application on the vehicle cluster to promote multi-tenancy, which improves resource efficiency but also makes it harder to find a solution.

\section{Results}

\subsection{Application Types}

\hl{We highlight two different application types that are suitable for the model described in this paper.} The \hl{type-based} service for distributed video analytics is given as an input to the service placement problem. The service is described as a linear chain, with one \hl{task} of \emph{Type 1} type, that is mapped to a vehicle with a dash camera or smart camera installed on it and the user is willing to lease their vehicle resources in exchange for some incentive. This data is streamed to a nearby vehicle that hosts a \hl{task} of \emph{Type 2}, followed by another vehicle hosting a \hl{task of} \emph{Type 3}. Such tasks execute lightweight video pre-processing like data compression or sub-sampling that reduces the size of the video data, based on the application requirement. Some examples include:
\begin{itemize}
    \item Modality based pre-processing \cite{8579129}: multimedia data may have more than one modality, e.g., video data with image and speech. This requires data separation.
    \item Data cleaning: only frames that have the required data can be separated from other redundant frames, especially in the case of more than one source of video data. This is relevant for a Fog computing scenario, where the computation and storage capacity is limited. 
    \item Data Reliability: Other application like detecting video from unreliable data sources which are not subscribed to the service can also be detected and filtered at this stage. 
\end{itemize}

 Once the processing is complete at the cluster, this data is then sent to the CN which forwards the data to the edge/cloud for further high computational processing, like vision-based processing for video crowd-sourcing applications and traffic density estimation using convolutional neural networks, etc. We specify two different applications, with different resource requirements, that we place together on the vehicle cluster:

\subsubsection{Application I: High-processing video streaming applications}

\begin{figure}[t!]\centering
\includegraphics[width=8cm]{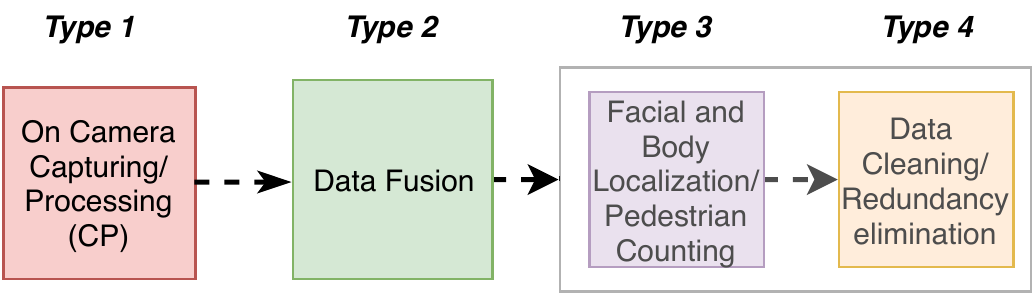}
\setlength{\belowcaptionskip}{-10pt}
\caption{Pedestrian Detection service concepts}
\label{fig:ServiceModel}
\end{figure}
The first application is a \emph{pedestrian detection application} that can be used to study the popularity of a coffee shop or a gas station, based on the number of pedestrians detected in the stream of video data. This data is collected by vehicles standing at a traffic light or an intersection, close to the coffee shop, say. This data has local relevance/scope and hence, most of this data should be processed locally, based on the available resources on the vehicle cluster. Such applications can be considered compute-intensive and require higher processing capacity and so large amounts of unnecessary data should not be uploaded to the cloud before processing. For this application, 1 to 6 camera or \emph{Type 1} \hl{TIs} are used in the Evaluation section (\ref{sec:Results}), because more camera instances increase the richness of context data. The \emph{Type 2} \hl{TIs} aggregate and process this video stream from different \emph{Type 1} \hl{TIs} as depicted in the Concept Diagram of the application in Fig.~\ref{fig:ServiceModel}. This \hl{TI} can aggregate the data from different sources based on content or location similarity. The functionality of \emph{Type 3} and \emph{Type 4} \hl{TIs} is application-specific. In the compute-intensive application, lightweight video processing is performed on the video stream to transform it into other forms, e.g., capturing specific frames with license plates, or highlighting pedestrians or other objects of interest in each scene. We assume that the data is reduced to  40-50\% of its size, by \emph{Type 2} instances and to 20\% of its original size after processing by \emph{Type 3} \hl{TIs}. This pre-processed data is then sent to the CN, which forwards it to the RSU.

\begin{figure}[tb]
\begin{minipage}[t]{\linewidth}
\begin{center}
    \includegraphics[width=\linewidth,height=5cm,keepaspectratio]{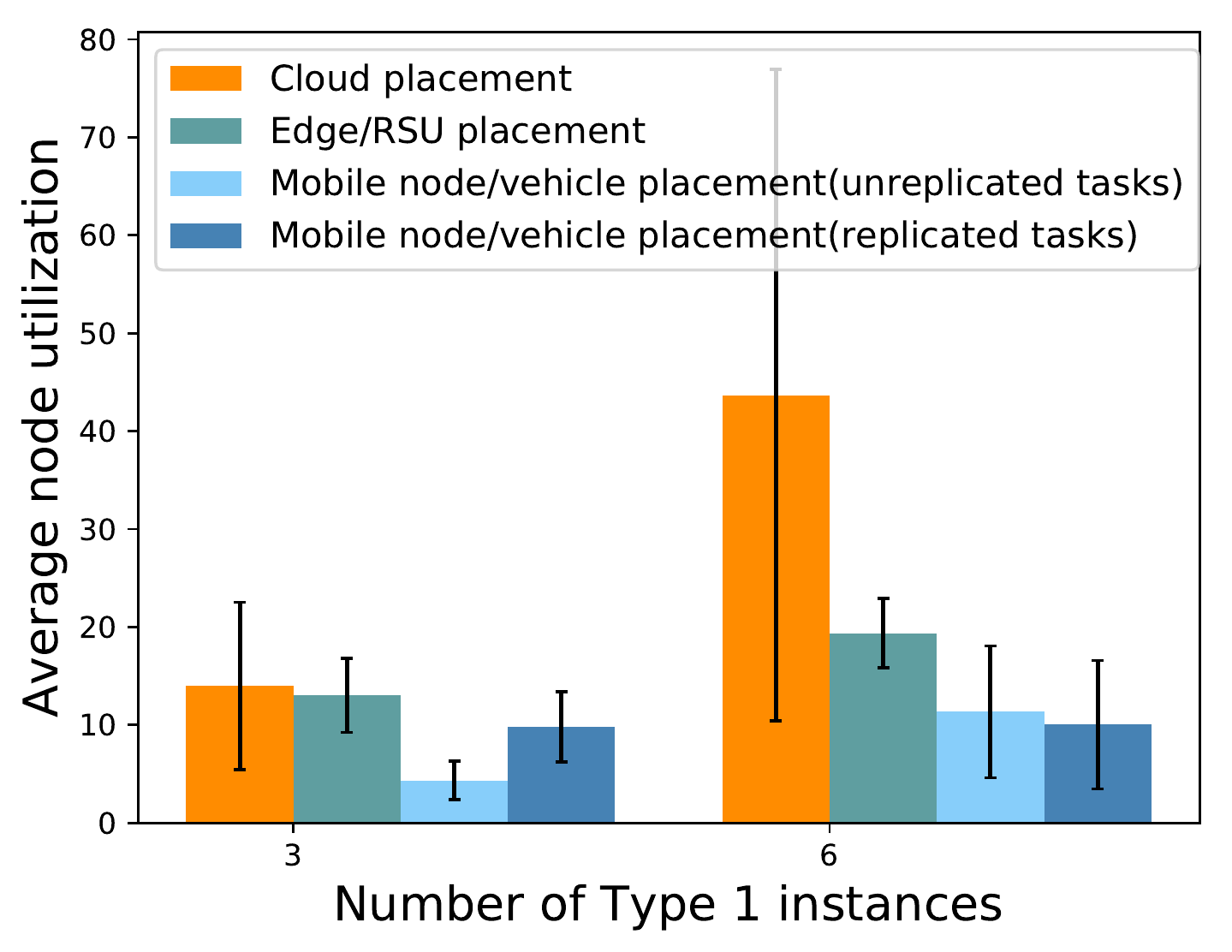}
    \end{center}
    \caption{Comparing different placement techniques using the average node utilization in the case of: cloud placement, edge/RSU placement, mobile node/vehicle placement (without replicating task instances), mobile node/vehicle placement (with replicated task instances: \textbf{our case}) for Application I 
    }
    \label{fig:YAFS}
\end{minipage}
\end{figure}

\hl{To validate the service model, we implement this applications on an existing simulator called Yet Another Fog Simulator (YAFS)\cite{8758823}.} It is a python-based discrete-event simulator that supports resource allocation and network design in Cloud, Fog or Edge Computing systems. We chose the simulator because it supports mobility of entities, which can act as both sources of data, called workloads or processing nodes. The simulator also provides a Distributed Data Flow based application model that allows \hl{task} replicas and dynamic placement of \hl{tasks}. The applications are represented as directed acyclic graphs (DAGs), where nodes represent service modules and links represent data dependency between modules. The simulator also incorporates strategies for dynamic service selection, placement and routing.


\hl{In Fig.~\ref{fig:YAFS}, we consider the average node utilization in the placement of service described as Application I. As suggested by the authors in \cite{8758823}, we calculate the node utilization as the sum of the service times at each node divided by the total simulation time. We compare the average node utilization between:}
\begin{itemize}
    \item \hl{Cloud placement: all task placed in Cloud}
    \item \hl{Edge/RSU placement: all tasks placed on edge/RSU}
    \item \hl{Mobile node/vehicle placement (unreplicated tasks): all tasks placed on mobile nodes/vehicles (without replicated TIs)}
    \item \hl{Mobile node/vehicle placement (replicated tasks) (our approach): all tasks placed as multiple TIs on mobile nodes/vehicle}
\end{itemize}
\hl{In \mbox{Fig.~\ref{fig:YAFS}}, for variable workloads that are generated using custom temporal distributions, we compare placement for three and six video collection TIs of Type 1. For three Type 1 TIs, only mobile node placement with unreplicated tasks result in better node utilization, compared to our approach. For six Type 1 TIs, our approach of replicating processing TIs of Type 2 and Type 3 on different mobile nodes, results in lesser average node utilization compared to all the other approaches. This validates that our service model of replicating tasks is efficient from node utilization point-of-view, as compared to other placement approaches.}


\subsubsection{Application II: Low-processing video streaming application}

Application II uses vehicles as moving sensors for video collection. Applications of this category includes measuring the traffic density at an intersection in real-time, or surveying road conditions for road traffic mapping. Generally, the focus is on passive video collection; most processing does not happen in the cluster. Such applications perform minor pre-processing tasks on data in the vehicle cluster. Such pre-processing includes data sampling, segmentation or encoding and is carried out on \emph{Type 2} and \emph{Type 3} instances. Thus, the data is reduced to 80\% of its original size before being sent to the cloud for executing compute-intensive tasks, possibly applying complex machine learning to the data. \\




\begin{figure*}[tb]
    \centering 
\begin{subfigure}{0.3\textwidth}
  \includegraphics[width=\linewidth,height=4.5cm]{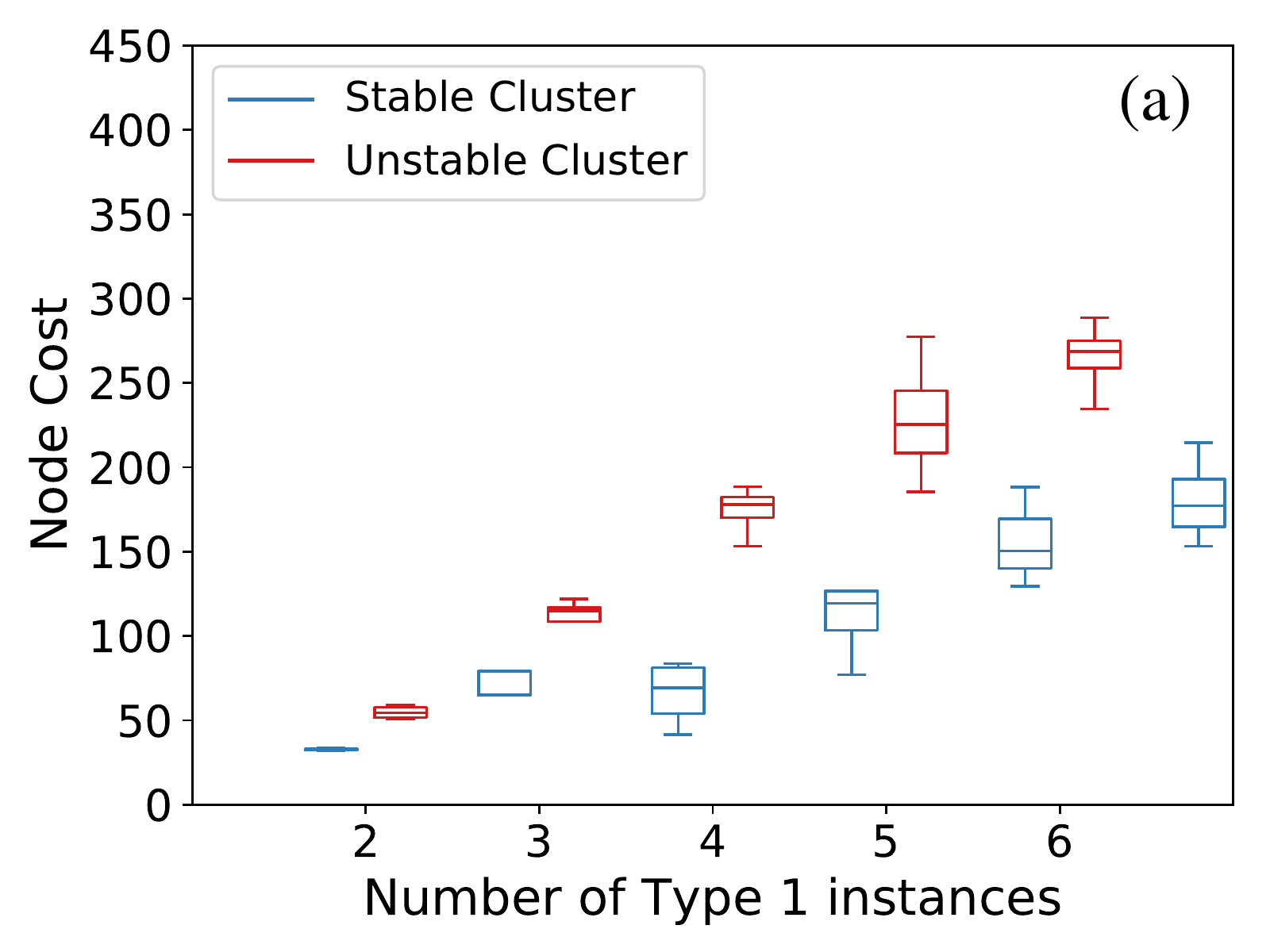}\vspace{-0.75em}
  \caption{\textbf{Case A:Node cost for resource-constrained cluster and low data rate}}
  \label{fig:NodeCost-ResourceConstrained-LowDataRate}
\end{subfigure}\hfil 
\begin{subfigure}{0.3\textwidth}
  \includegraphics[width=\linewidth,height=4.5cm]{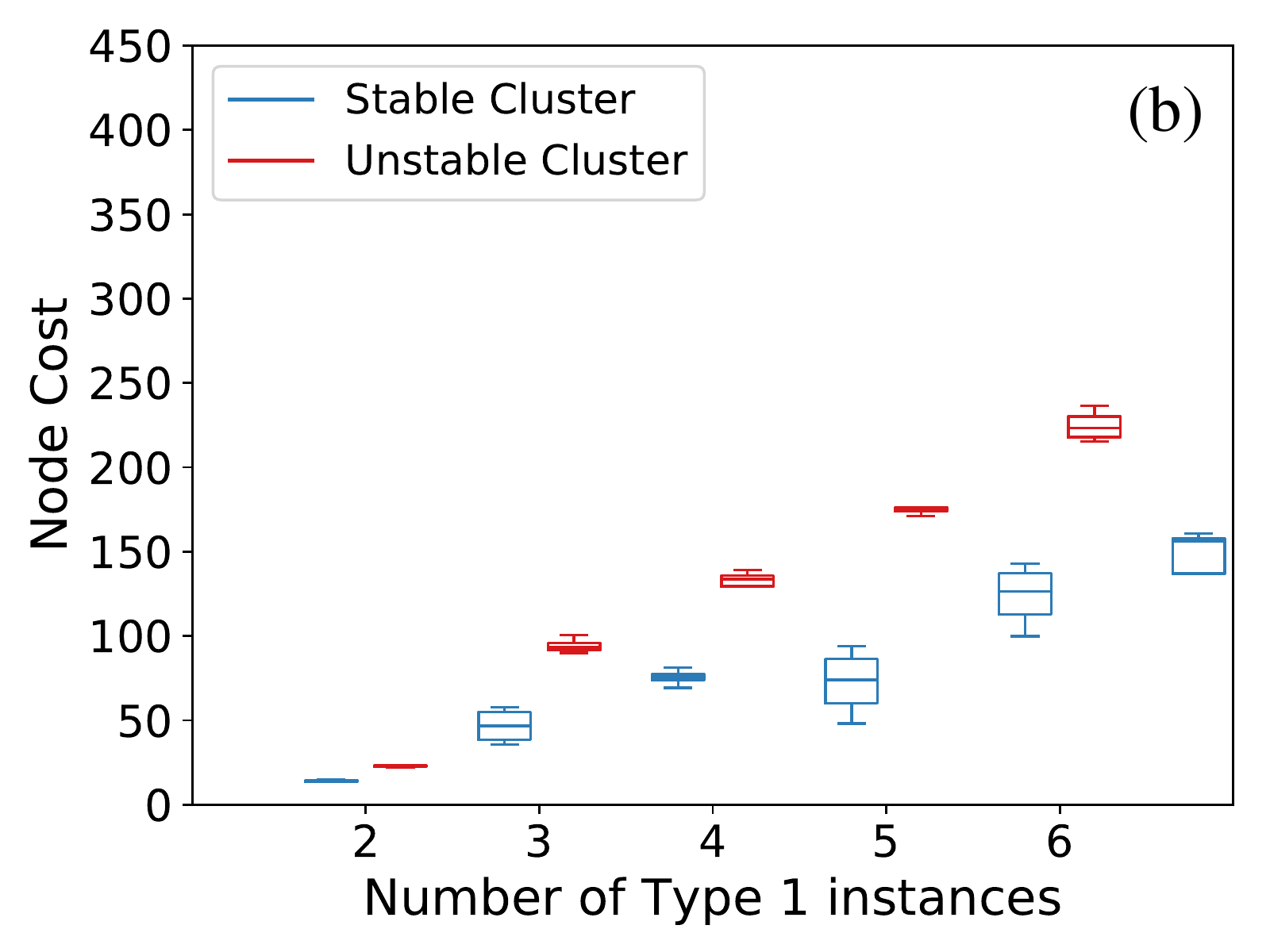}\vspace{-0.75em}
  \caption{\textbf{Case B:Node cost for resource-rich cluster and low data rate}}
  \label{fig:NodeCost-ResourceRich-LowDataRate}
\end{subfigure}\hfil 
\begin{subfigure}{0.3\textwidth}
  \includegraphics[width=\linewidth,height=4.5cm]{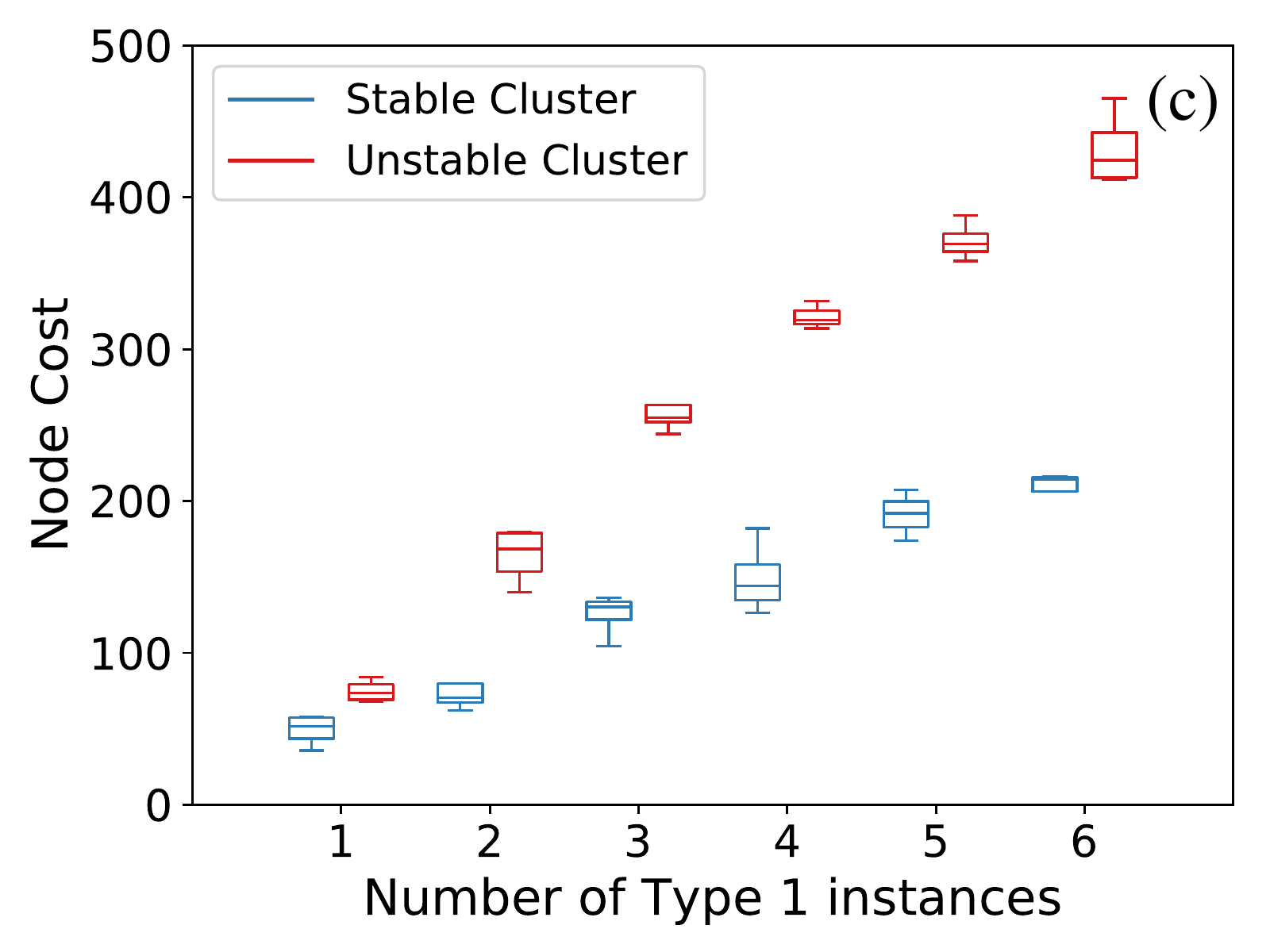}\vspace{-0.75em}
  \caption{\textbf{Case C:Node cost for resource-rich cluster and high data rate}}
  \label{fig:NodeCost-ResourceRich-HighDataRate}
\end{subfigure}

\medskip
\begin{subfigure}{0.3\textwidth}
  \includegraphics[width=\linewidth,height=4.5cm]{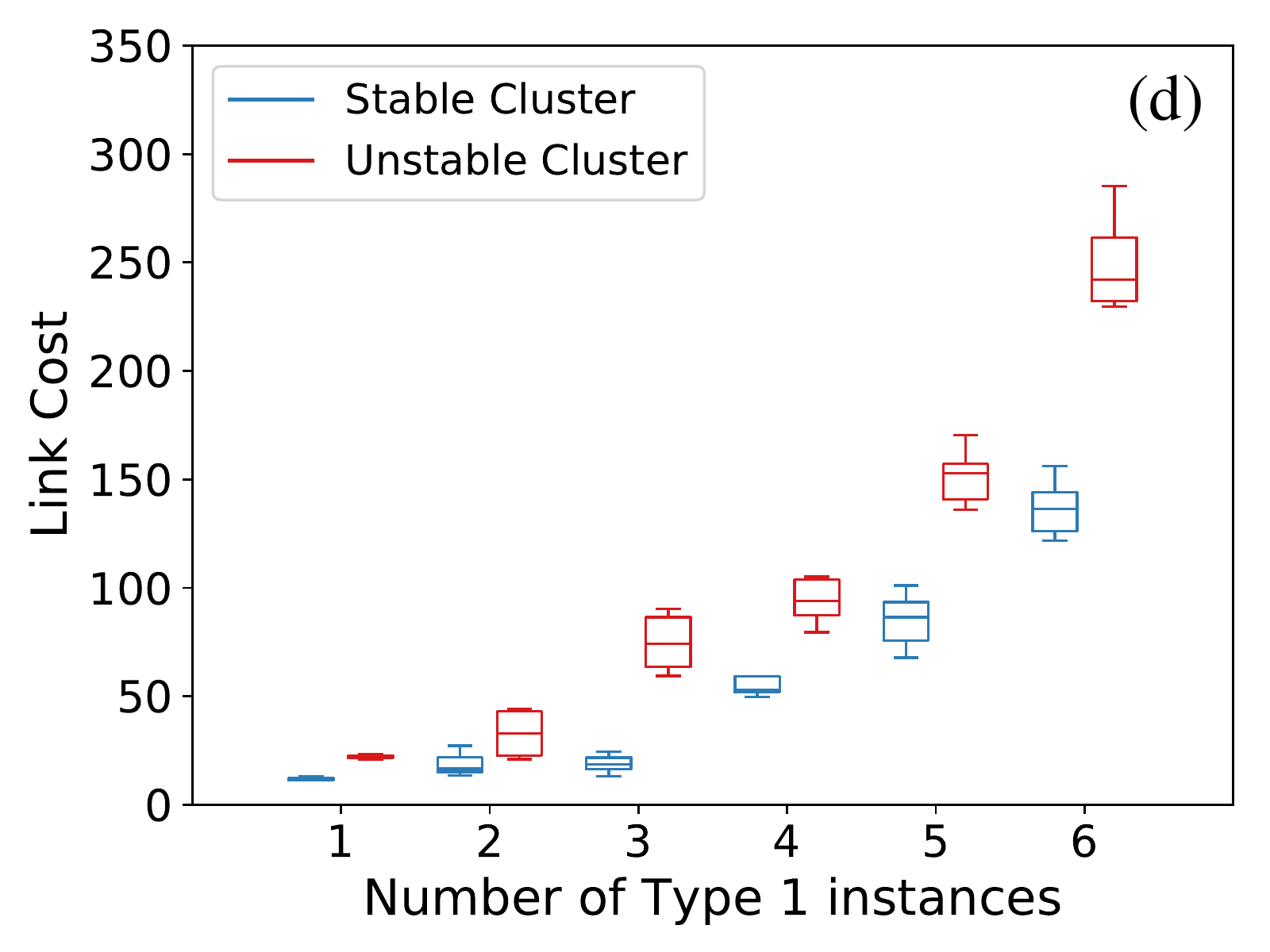}\vspace{-0.75em}
  \caption{\textbf{Case A:Link cost for resource-constrained cluster and low data rate}}
  \label{fig:LinkCost-ResourceConstrained-LowDataRate}
\end{subfigure}\hfil 
\begin{subfigure}{0.3\textwidth}
  \includegraphics[width=\linewidth,height=4.5cm]{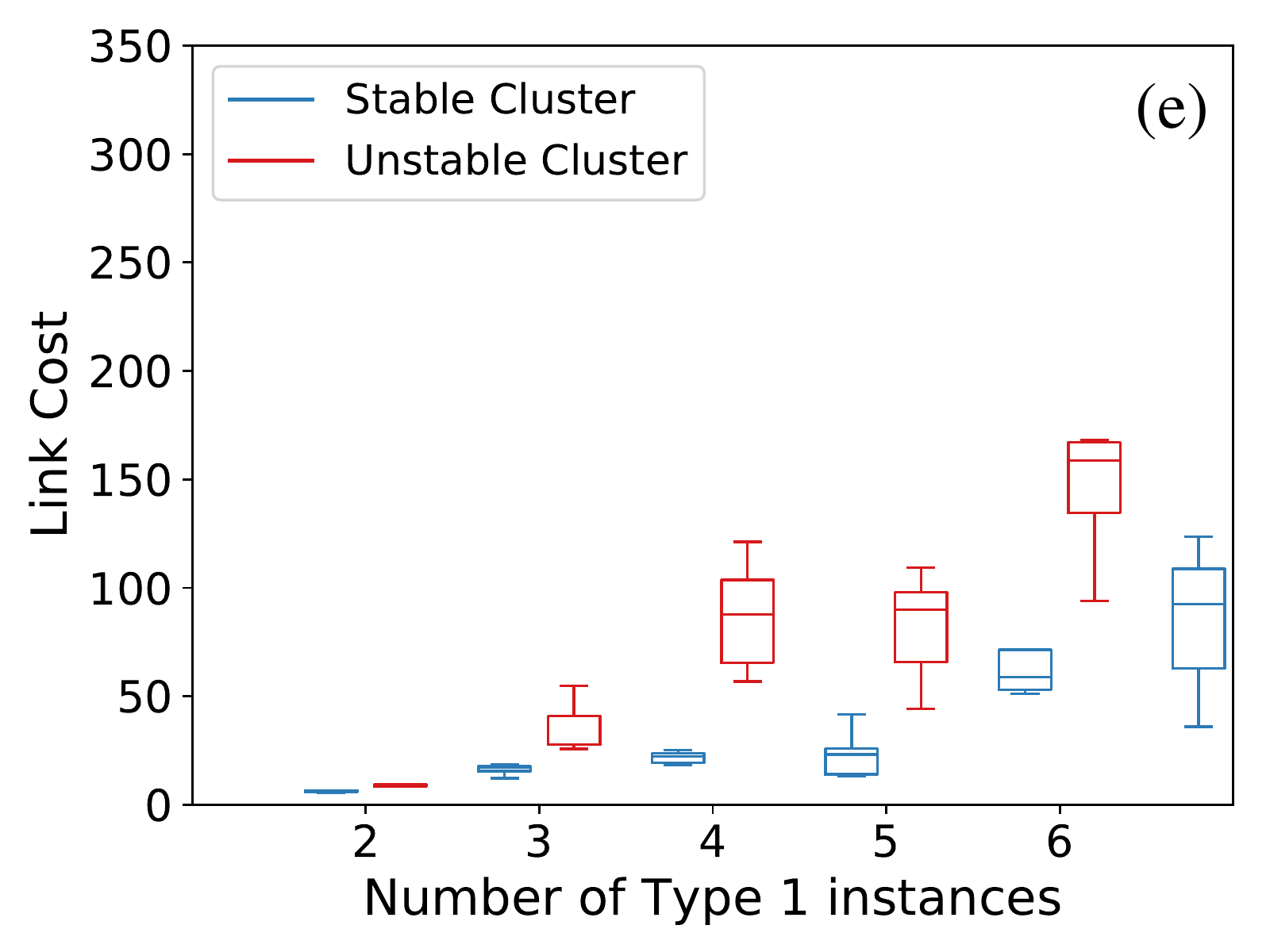}\vspace{-0.75em}
  \caption{\textbf{Case B:Link cost for resource-rich cluster and low data rate}}
  \label{fig:LinkCost-ResourceRich-LowDataRate}
\end{subfigure}\hfil 
\begin{subfigure}{0.3\textwidth}
  \includegraphics[width=\linewidth,height=4.5cm]{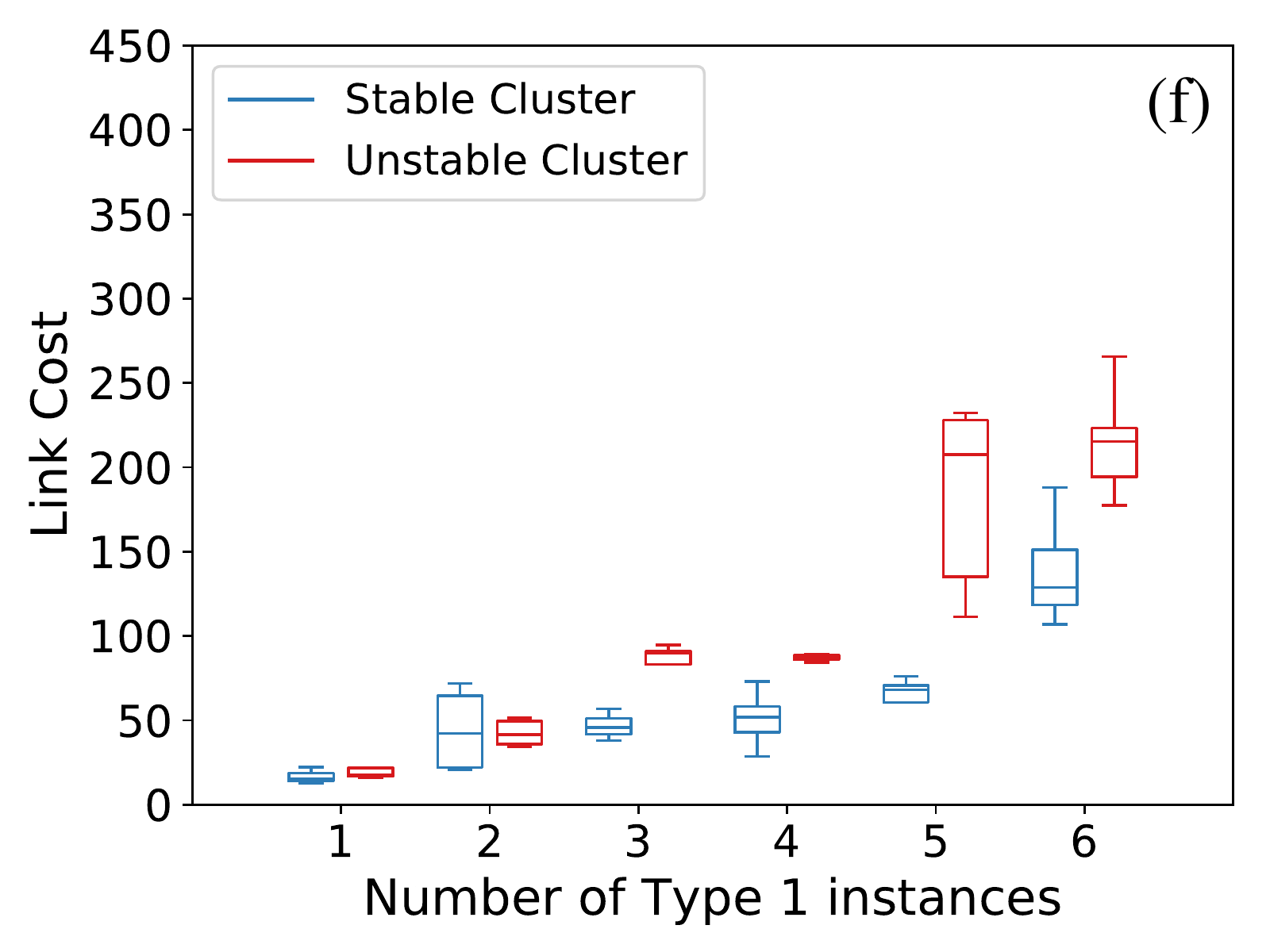}\vspace{-0.75em}
  \caption{\textbf{Case C:Link cost for resource-rich cluster and high data rate}}
  \label{fig:LinkCost-ResourceRich-HighDataRate}
\end{subfigure}
\caption{Node and Link Cost for Case A,B \& C}
\label{fig:images}
\end{figure*}
\subsection{Evaluation}
\label{sec:Results}


We solve the constrained optimization problem using the Gurobi solver, which is a powerful mathematical sover, on an Intel i7-6500U dual-core processor running at 2.50 GHz. The solver uses a Linear Programming (LP) based branch and bound algorithm to solve the Mixed Integer Programming (MIP) problem. 


We place Applications I and II together on a vehicle cluster with 10 nodes, since more nodes in a cluster increases the time and space complexity of the problem. The cluster is a directed, connected graph, where each node has either video capturing or data processing functionality. We consider two types of resource states of the cluster, based on the mix of vehicles with one of three resource profiles: 1) Large node type: 5 CPUs, 500Mb disk, 6MB/s bandwidth; 2) Medium node type: 3 CPUs, 250Mb disk, 4MB/s bandwidth; and 3) Small node type: 2 CPUs, 100Mb disk, 2MB/s bandwidth. A \emph{resource-rich cluster} has 50\% large, 25\% medium and 25\% small resource vehicle nodes. A \emph{resource-poor cluster} has 25\% large, 50\% medium and 25\% small vehicle nodes.
We consider a service chain with 2 processing instances, which makes the chain length = 3, including Type 1 instances and the CN. We ran the optimization for the longer chain length, which takes a much longer time to find a solution, specially for a higher number of video generating instances, with higher data rate. The worst case scenario was for a service chain of length 6 with 5 Type 1 instances, which took more than 5 hours to find a solution. 

The 'type graph' is scaled as an 'instance graph', with data dependency and resource requirements. We use a service chain description similar to~\cite{8459915}, without making it bidirectional. We impose multi-tenancy in the model, as it is beneficial to share \hl{TIs} between applications, especially when more than one \hl{task} replica is placed on the vehicle cluster.

For this paper, we consider that all nodes stop at an intersection and the RSU first selects a CN, which is one hop away from the RSU and is well connected to more than 70-80\% of the nodes in the cluster. This CN needs to have ample communication and computation resources to manage the resource and cluster state. We also assume that the mobility behavior, in terms of the CCP is based on the mobility pattern of each vehicle, collected over its previous trips in this area. We derive the CCP by running the calibrated SUMO simulator, using the real vehicle density data from Dublin traffic, as explained in  \S\ref{sec:CCP}. We have broadly classified cluster states as stable and unstable. The stable clusters are formed when many vehicles follow a single trajectory, along with the CN. We consider two cluster states: \emph{stable} with a CCP in the range [0.4,0.8] and \emph{unstable} with a probability distribution between [0.2,0.6]. 



We consider three use cases for solving the optimization. For Case A, we take a resource-constrained cluster with low data rates of streaming video, and compare the node processing cost for stable and unstable cluster probabilities. We vary the number of Type 1 instances from 2 to 6, to study the effect of the amount of data on service placement and resource usage. When we use lower video data rates, we see that the stable cluster uses less resources than the unstable cluster. For Case B, resource-rich case (Fig.~\ref{fig:NodeCost-ResourceRich-LowDataRate}) with lower data rates, the node cost is significantly less, compared to Case A (Fig.~\ref{fig:NodeCost-ResourceConstrained-LowDataRate}), as it is easier to place more than one \hl{TI} on nodes having more processing resources, for both stable and unstable cluster, resulting in better resource utilization. But in this case, the stable cluster still used less resource than the unstable cluster. The solution time is also significantly less for a resource-rich cluster: to find the optimal placement for Case B takes an average of 86s, versus a resource-constrained cluster (Case A: 300.7s). We also observed that weighting both objectives (adjacency TI placement and total cost of service placement) equally solves the problem faster than hierarchical solving, but the resulting placement uses more network resources.

For link cost, the resource-constrained cluster (Case A) has significantly higher resource usage (Fig.~\ref{fig:LinkCost-ResourceConstrained-LowDataRate}). The nearby nodes might not have enough processing capacity, so dependent \hl{TIs} need to be placed on farther nodes, leading to more link utilization. The link capacity is also less in the resource-constrained cluster which adds to the cost. The Link Cost in the resource-rich cluster (Case B) (Fig.~\ref{fig:LinkCost-ResourceRich-LowDataRate}) is significantly less and, in both cases, stable clusters outperform unstable clusters. The variability in link cost is more in this case, as the amount of video data processing in both applications is significantly different. Application I reduces the data to approx. 20\% whereas Application II reduces the data to 80\%. Hence, the link cost varies based on the number of \emph{Type 1} \hl{TIs} in each application. But as we double the data rate of the video data in a resource-rich cluster (Case C), the unstable cluster utilizes much more computation resource (Fig.~\ref{fig:NodeCost-ResourceRich-HighDataRate}). The difference between stable and unstable cluster node cost increases significantly as the Type 1 instances increase from 1 to 6. The unstable cluster uses slightly more resources, compared to the stable cluster in the low data rate case (Case B: Fig.~\ref{fig:NodeCost-ResourceRich-LowDataRate}). For the link cost in this case (Fig.~\ref{fig:LinkCost-ResourceRich-HighDataRate}), for fewer Type 1 TIs, stable and unstable cluster incur almost the same cost. The variability increases as the number of video instances increases. 

\begin{figure}[tb]
\begin{minipage}[t]{\linewidth}
\begin{center}
    \includegraphics[width=\linewidth,height=5cm,keepaspectratio]{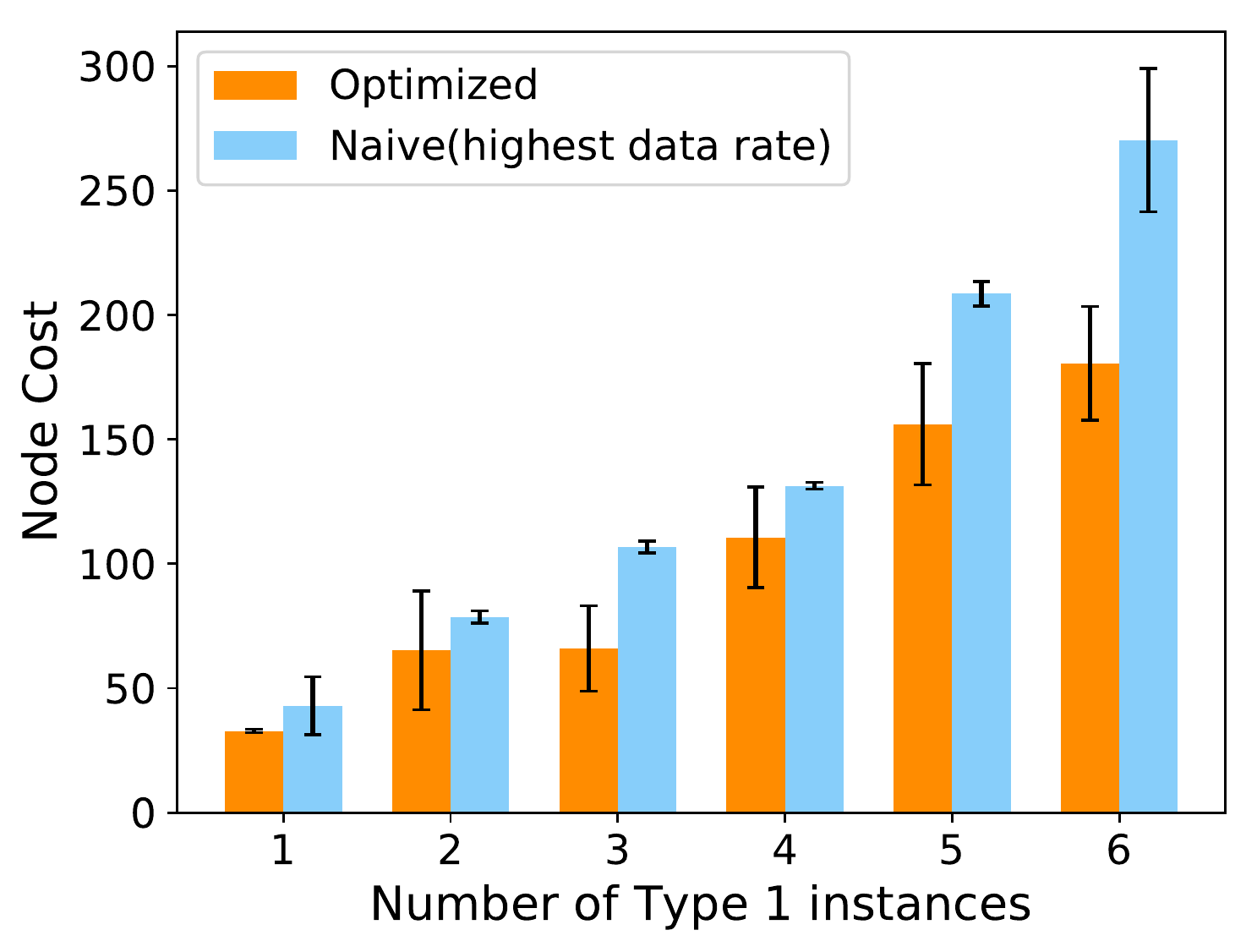}
    \end{center}
    \caption{Comparison of node cost in optimal v/s naive scheme for Case A:resource constrained cluster and low data rate}
    \label{fig:f7}
\begin{center}
    \includegraphics[width=\linewidth,height=5cm,keepaspectratio]{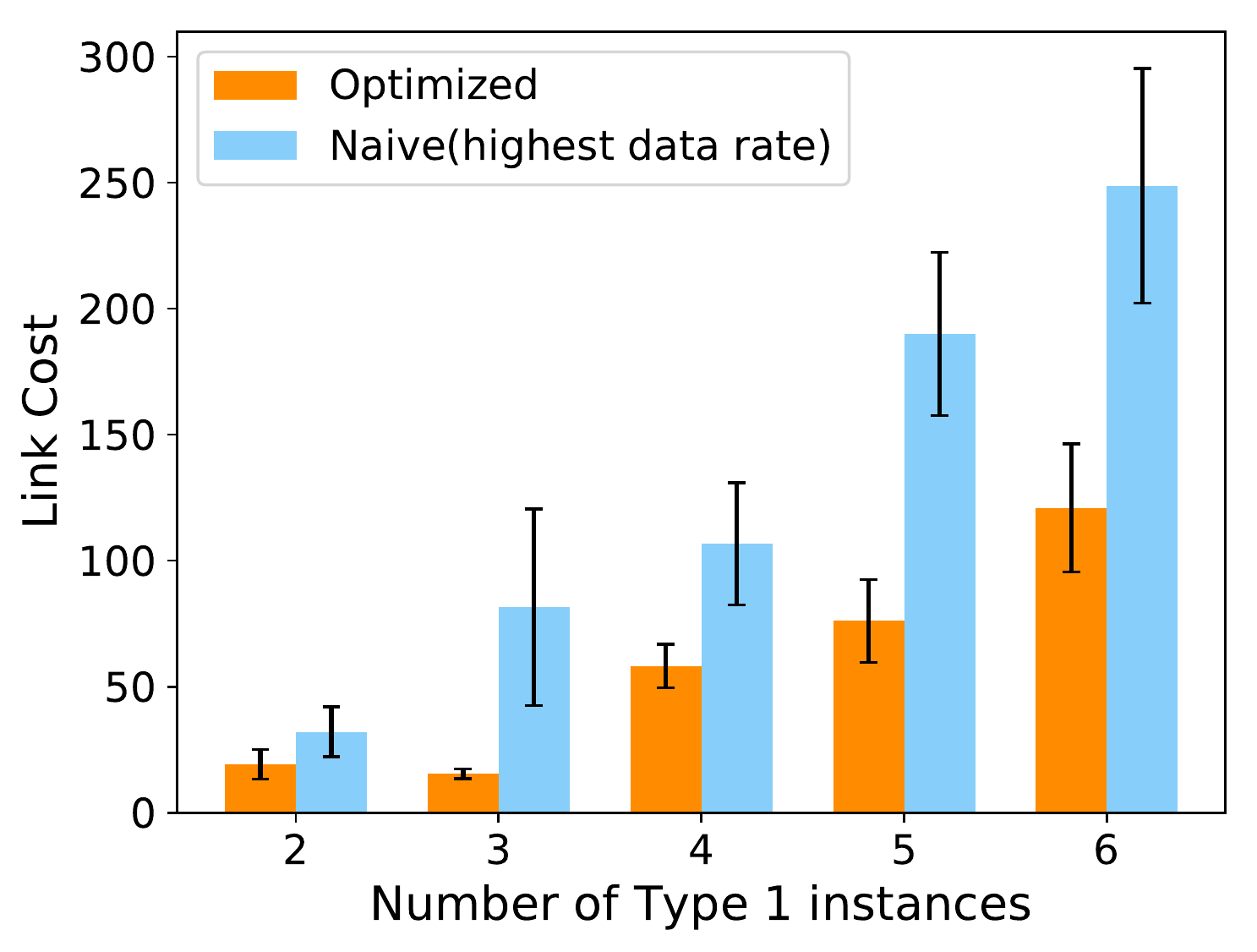}
    \end{center}
    \caption{Comparison of link cost in optimal v/s naive scheme for Case A:resource constrained cluster and low data rate}
    \label{fig:f8}
\end{minipage}
\end{figure}

We compare the optimal solution \hl{of our model} to a naive solution introduced in \cite{7946184}. The Autonomous Vehicular Edge Computing + Naive solution selects combinations of nodes with the intention of minimising latency, i.e., they have the lowest processing time and transmission time. As we focus on data collection services, we do not focus on the time needed to send results to the requesting node (typically the CN). As described in \cite{8489874}, they preferentially select nodes with the highest available link and node capacities. As seen in Fig.~\ref{fig:f7}, the node cost for the naive approach increases significantly as the number of Type 1 instances increases. We get a similar result for link cost in Fig.~\ref{fig:f8} where the cost doubles for 5 and 6 Type 1 TIs, in comparison to the optimal solution. The naive approach results in less latency compared to the optimal approach, but does not take account of node mobility, so is more likely to fail. By contrast, our objective reduces the cost of service execution \emph{and} selects more reliable nodes that reduce the need for service reconfiguration. Of course, delay is a crucial parameter for safety-related services like lane changing, accident prevention and autonomous driving. However, delay can also be reduced by adding more resources and using them judiciously: by shortening service chains and placing many processing \hl{TIs} of the same \hl{type} in parallel.

\subsection{Penalty Function} We introduce a Penalty function to reflect the cost of nodes hosting \hl{TIs} leaving the cluster. This penalty is related to both link and node cost. If a \hl{TI}-hosting node leaves the cluster, the node cost incurred by it does not contribute to the service requirement. In case of service reconfiguration, the leaving node will have to send the service state back to the CN, which requires bandwidth usage. We penalize placements where \hl{TIs} are placed on nodes with a probability lower than the threshold probability (0.6). There will be another added cost on re-configuring the failing service, but we do not consider service reconfiguration cost in our current model.

\begin{figure}[tb]
\begin{subfigure}{0.5\linewidth}
  \includegraphics[width=\linewidth,height=3cm,keepaspectratio]{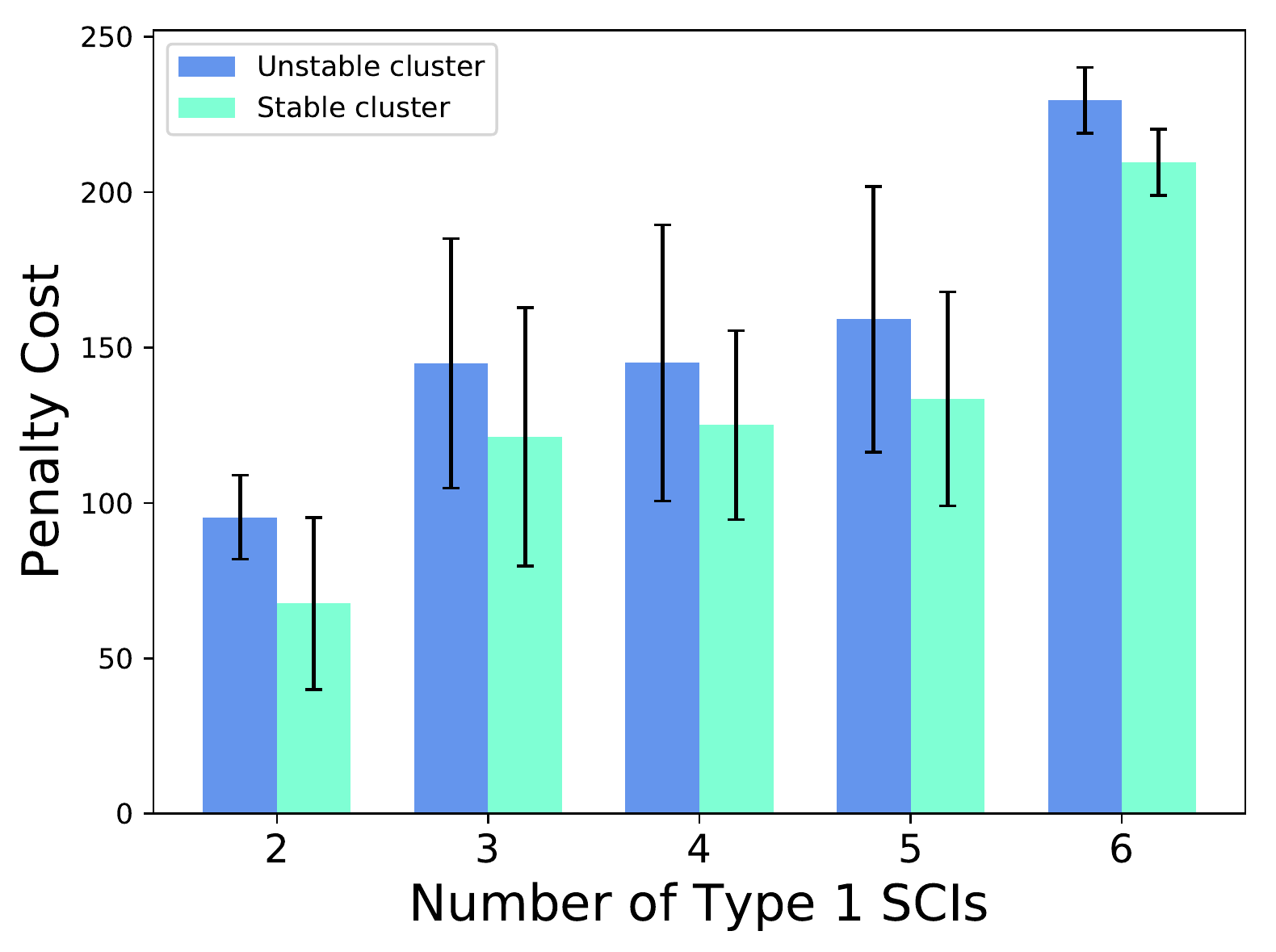}
  \caption{Penalty Cost}
  \label{fig:PenaltyCost-ResourceConstrained}
\end{subfigure}\hfil 
\begin{subfigure}{0.5\linewidth}
  \includegraphics[width=\linewidth,height=3cm,keepaspectratio]{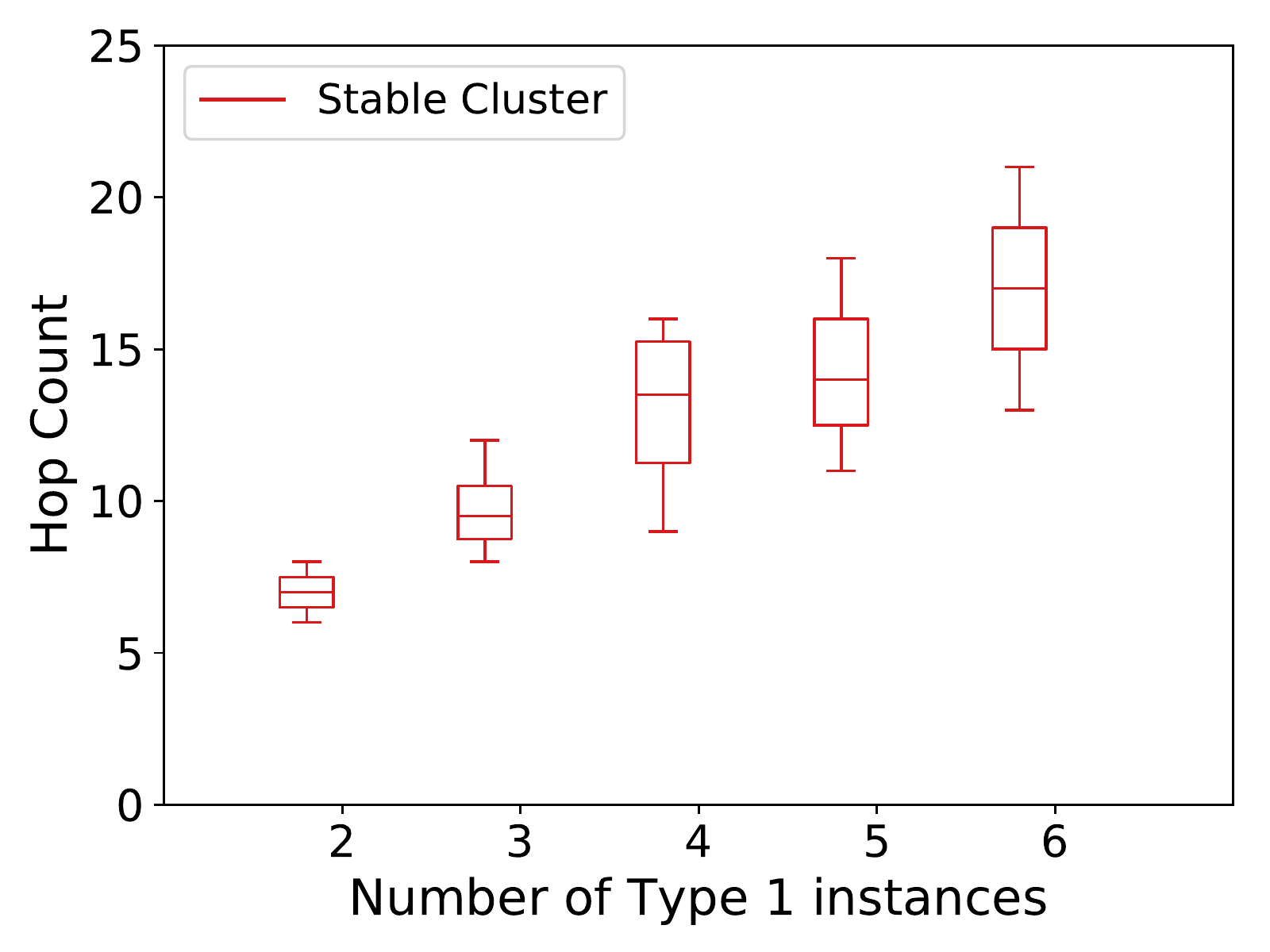}
  \caption{Total Hop Count}
  \label{fig:TotalHopCount-ResourceConstrained}
\end{subfigure}\hfil 
\caption{Penalty and Hop Count in resource-constrained cluster}
\end{figure}

The Penalty function for selecting two nodes with a joint probability (to stay with the cluster) less than the threshold probability is given as:
\begin{equation}
\small{
P(\lambda_{(i_1,i_2)}) =  \lambda(P_{(t_1,t_2)}(i_1,i_2) - P_{Threshold})  
}
\label{eq:LinkPenaltyFunction}
\end{equation}
where $\lambda$ can take values like 5,10,..,100, based on how strongly we want to penalize task placement on nodes with lower CCP. As shown in Fig.~\ref{fig:PenaltyCost-ResourceConstrained}, for a resource constrained cluster (Case A), the penalty cost added to the objective function for an unstable cluster is higher than the penalty for a stable cluster. For the resource-rich cluster (Case B), the model places very few TIs on nodes with low CCP, and hence the additional penalty cost is zero, in almost all the resource-rich test cases, for both stable and unstable clusters. This suggests that the model considers mobility in an intuitive way. We also plot the best and worst case hop count (Fig.~\ref{fig:TotalHopCount-ResourceConstrained}), for the same resource-constrained cluster (Case A), when the total hop count is first minimized for service selection. It is observed that the hops increase significantly as the number of Type 1 instances increase. 

\subsection{Mininet-WiFi Simulation}
\begin{figure}[tb]
\begin{minipage}[t]{\linewidth}
\begin{center}
    \includegraphics[width=\linewidth,height=5cm,keepaspectratio]{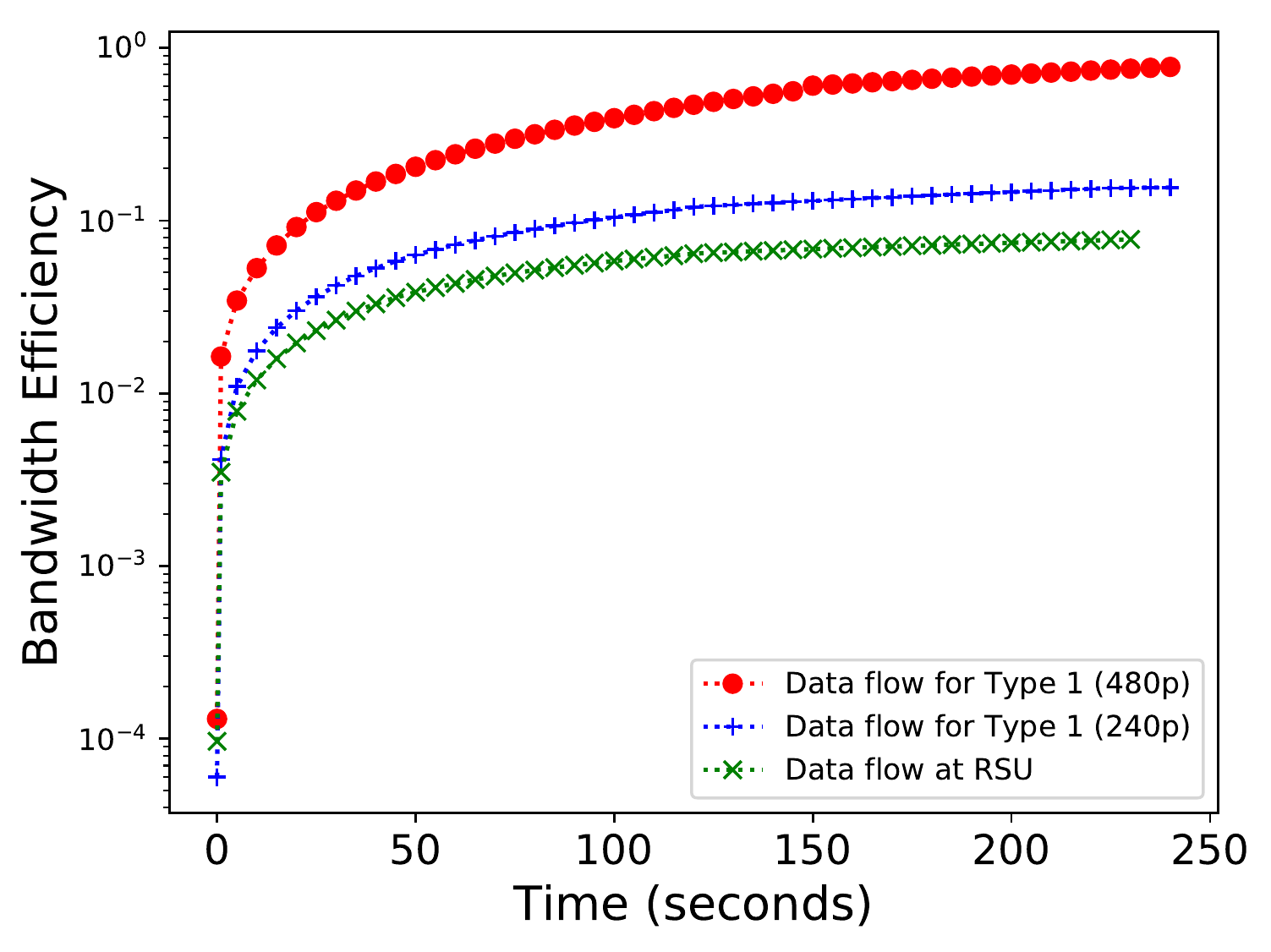}
    \end{center}
    \caption{Bandwidth Efficiency at Type 1 TIs and RSU}
    \label{fig:f2}
\end{minipage}
\end{figure}
We emulate the scenario of a vehicle cluster with 10 nodes for Application II. We simulate two video streaming nodes that send 4 minutes long video in different resolutions (240p and 480p) and at different data rates. We use a SUMO-Mininet-WiFi \cite{7859348} set-up, where the RSU receives a service monitoring request at the same Dublin intersection and forwards the request to the CN. We use the SUMO simulator to generate urban mobility at a busy intersection. Mininet-WiFi maps the cars to an emulated, software-defined network with virtualized WiFi station and access points. The video stream is forwarded to a participating node that aggregates the data before sending it to a CN that forwards it to the RSU. We monitor the bandwidth usage for the two collected streams and compare it to the bandwidth efficiency of the processed stream received at the RSU. As depicted in Fig.~\ref{fig:f2}, we get significantly better bandwidth efficiency at the RSU because of the data aggregation in the model.

\section{Conclusion and Future Work}

This paper focuses on the concept of scaling and placement of distributed services on vehicle clusters, harnessing the knowledge of mobility patterns. The novelty of the work is in considering urban road traffic as a potential site for deploying services. The services can be scaled dynamically, based on the resource and mobility state of the multi-hop cluster. We introduce a flow model for the traffic, study predictability in vehicular flow, and estimate communication capacity using real vehicular traffic data. We introduce a detailed mathematical model for the mobility-aware scaling of distributed services based on resource-rich and resource-poor as well as stable and unstable cluster states. We solve the constrained bi-objective optimization problem, introduce data collection and data pre-processing applications, and validate our model for different resource and mobility states. Our approach outperforms the naive solution introduced in \cite{7946184} significantly. 

As part of the future work, a decentralized, mobility-aware task offloading algorithm will be introduced that solves the optimization problem in real-time. To make the service model more practical, we will introduce a distributed service reconfiguration scheme to send collected data or service state back to the vehicle cluster. We aim to use hyper-parameter optimization techniques to decide the number of TIs to be deployed in real-time, for satisfying the service placement requirements. We will focus on the replacement of concurrent, data-dependent tasks as part of the failure recovery scheme. 


\bibliographystyle{IEEEtran}
\IEEEtriggeratref{28}

\bibliography{access}


\end{document}